\documentclass[12pt,preprint]{aastex}
\usepackage{epsf}
\usepackage{natbib}
\usepackage{float}
\usepackage{appendix}
\usepackage{amsmath}
\usepackage{pdflscape}

\tabletypesize{\scriptsize}

\def\gappeq{\mathrel{ \rlap{\raise.5ex\hbox{$>$}}
                      {\lower.5ex\hbox{$\sim$}}  } }

\begin{document}
\slugcomment{Accepted for Publication in ApJ}
\shorttitle{Maximizing ExoEarth Yield}
\shortauthors{}

\title{Maximizing the ExoEarth Candidate Yield from a Future Direct Imaging Mission}

\author{Christopher C. Stark\altaffilmark{1}, Aki Roberge\altaffilmark{2}, Avi Mandell\altaffilmark{2}, Tyler D. Robinson\altaffilmark{3}}

\altaffiltext{1}{NASA Goddard Space Flight Center, Exoplanets \& Stellar Astrophysics Laboratory, Code 667, Greenbelt, MD 20771; christopher.c.stark@nasa.gov}
\altaffiltext{2}{NASA Goddard Space Flight Center, Greenbelt, MD 20771}
\altaffiltext{3}{NASA Ames Research Center, Moffett Field, CA 94035}

\begin{abstract}

ExoEarth yield is a critical science metric for future exoplanet imaging missions.  Here we estimate exoEarth candidate  yield using single visit completeness for a variety of mission design and astrophysical parameters.  We review the methods used in previous yield calculations and show that the method choice can significantly impact yield estimates as well as how the yield responds to mission parameters.  We introduce a method, called Altruistic Yield Optimization, that optimizes the target list and exposure times to maximize mission yield, adapts maximally to changes in mission parameters, and increases exoEarth candidate yield by up to 100\% compared to previous methods.  We use Altruistic Yield Optimization to estimate exoEarth candidate yield for a large suite of mission and astrophysical parameters using single visit completeness.  We find that exoEarth candidate yield is most sensitive to telescope diameter, followed by coronagraph inner working angle, followed by coronagraph contrast, and finally coronagraph contrast noise floor.  We find a surprisingly weak dependence of exoEarth candidate yield on exozodi level.  Additionally, we provide a quantitative approach to defining a yield goal for future exoEarth-imaging missions.

\end{abstract}

\keywords{telescopes --- methods: numerical --- planetary systems}

\section{Introduction}
\label{intro}

The number of habitable extrasolar Earth-like planets (exoEarths) detected and spectroscopically characterized is a key scientific metric for future missions that aim to find habitable conditions on exoplanets and possibly signs of life.  The predicted exoEarth yield will inform telescope design and mission lifetime, and may ultimately help select high-contrast imaging technologies.  Radial velocity measurements and results from the \emph{Kepler} Mission have constrained the fraction of M dwarfs that harbor roughly Earth-sized planets in their habitable zones (exoEarth candidates), $\eta_{\Earth} \sim 0.4$ \citep{2013A&A...549A.109B,2013ApJ...767L...8K}, although the uncertainties of these estimates are large and the planet radius/mass and orbital period range used to define $\eta_{\Earth}$ vary within the literature.  For Sun-like stars $\eta_{\earth}$ is even less certain, with estimates ranging from $\sim$ 0.05 -- 0.2 \citep{2013PNAS..11019273P,2014arXiv1404.2960M,Silburt:2014aa}.  For such low values of $\eta_{\earth}$ the expected total number of candidate exoEarths observed within the mission lifetime (exoEarth candidate yield) could be low, making yield maximization extremely important.

Many previous studies have estimated the exoEarth candidate yield for future exoplanet imaging missions.  \citet{2004ApJ...607.1003B} introduced the concept of obscurational completeness to quantify the impact of a non-zero inner working angle (IWA) on exoplanet detection.  In a landmark paper, \citet{2005ApJ...624.1010B} then built on this concept to include photometric completeness in combination with exposure time estimation to prioritize a target list and estimate exoEarth candidate yield for a coronagraphic mission.  Other studies have followed, with models varying in complexity and realism \citep[e.g.,][]{2010ApJ...715..122B,2010PASP..122..401S,2012PASP..124..418T}.  Most of these models seek to maximize the chance of observing an exoEarth around a given target and prioritize the target list based upon a cost to benefit ratio, but few have sought to optimize observations to maximize overall mission yield \citep{2007SPIE.6693E..22H}.

Here we introduce the Altruistic Yield Optimization (AYO) method for target prioritization and exposure time allotment that has the single goal of maximizing the total yield of exoEarth candidates.  In Section \ref{caveats} we discuss the major assumptions and limitations of this work, then provide a detailed look at the astrophysical assumptions in Section \ref{astrophysical_assumptions_section}.  Section \ref{exoEarth_yield_section} presents a review of yield calculation methods as well as our AYO method.  In Section \ref{results_section}, we compare the yield using AYO to previous methods and use AYO to estimate the yield as a function of several mission parameters.  We discuss and summarize our findings in Sections \ref{discussion} and \ref{conclusions}.

\section{Framework \& Caveats}
\label{caveats}

In this work we estimate exoEarth candidate yield using simple completeness and exposure time calculators, analogous to several previous studies \citep[e.g.,][]{2005ApJ...624.1010B,2012PASP..124..418T}.  Our exoEarth candidate yield calculations assume that all targets can be observed for the desired exposure time.  We ignore mission scheduling constraints and realtime cost-based decision making that are included in more complex ``mission execution simulators" \citep[e.g.,][]{2010PASP..122..401S}.  However, the yield maximization methods described below run rapidly ($\sim$ few seconds on a single $2.4$ GHz processor) and can be used to inform realtime decisions in mission execution simulators.

We make a number of assumptions and approximations to simplify our estimates of exoEarth candidate yield.  We budget one year of exposure time for all exoEarth observations.  We do not include overheads associated with the observations (e.g., telescope slew times, coronagraph wavefront correction time, etc.) within this budget.  Estimates for coronagraph overheads are poorly constrained at this time, as they depend strongly on many detailed mission design characteristics, including the coronagraph used, the telescope thermal and mechanical stability, and the wavefront control system used.  We therefore ignore overheads for now and simply note that the total exposure time budget of one year is not the total time that must be devoted to exoplanet detection.

The study presented here focuses solely on broadband detections; we do not include spectral characterization time for any of the detected exoEarth candidates.  The spectral characterization time depends on the bandpasses required for water and biosignature detection and the spectral resolving power, which in turn require well-defined mission science goals.  In addition, the spectral characterization time can depend strongly on the observational procedure, e.g., will low resolution spectra be obtained for every potential point source to help discriminate between planets and background objects, or will spectra only be obtained after common proper motion is established?  These complicating factors require additional analysis that is beyond the scope of this paper.  We note that in the case $\eta_{\earth} \lesssim 0.1$, as we will assume, the spectral characterization time is a modest fraction of the total time budget because the proportion of exoEarth candidate spectra obtained to stars imaged is relatively small, and in the case of single-visit studies, as discussed below, the yield scales weakly with mission lifetime.  Therefore, spectral characterization time should not greatly impact the single-visit yield estimates from this study.

Finally, we observe each star only once and ignore revisits in this study.  Revisits can both increase and decrease mission yield.  In the event that an exoEarth is not detected around a given star, revisiting that star at a later time, after any potential exoEarths have had a chance to move along their orbits, could result in a new detection.  On the other hand, revisits will also be required to establish common proper motion and constrain the orbit of any detected exoEarth candidates, reducing the amount of time available for new exoEarth candidate detections around other stars.  A proper calculation of how revisits impact mission yield requires optimizing the time between visits, which is work that we leave for future development.  Because we do not include revisits, we place no restriction on the maximum exposure time for single-visit detection.

The astrophysical parameters assumed for our baseline study are listed in Table \ref{exoearth_params_table}.  We choose $\eta_{\Earth} = 0.1$, consistent with recent estimates ranging from $\sim$ 0.05 -- 0.2 for Sun-like stars \citep{2013PNAS..11019273P,2014arXiv1404.2960M,Silburt:2014aa}.  We assume all exoEarth candidates are Earth twins, an assumption we address in Appendix \ref{earth_albedo_appendix}.  We use the same habitable zone conventions as \citet{2005ApJ...624.1010B}, distributing all planets uniformly and linearly in semi-major axis from $0.7$--$1.5$ AU (scaled with stellar luminosity to maintain constant HZ temperature) and in eccentricity from $0$--$0.35$.  The assumptions of this orbital distribution and the impact on yield will be addressed in a future paper.  We assume the fraction of stars with an Earth-sized planet in the habitable zone is $\eta_{\Earth}=0.1$.  Estimates of $\eta_{\Earth}$ include all habitable zone planets that are roughly Earth-sized---there is no guarantee that these planets are Earth-like.  Strictly speaking, we calculate a yield of \emph{candidate} exoEarths, a subset of which may be truly Earth-like.  

Table \ref{baseline_params_table} lists the baseline telescope and instrument parameters.  We use optimistic parameters for a coronagraph under the assumption of a direct imaging mission several decades in the future.  We ignore read noise and dark counts, implicitly requiring that future detector noise count rates be less than the astrophysical noise, and assume a diffraction-limited PSF.

We require a signal to noise ratio (S/N) of 10 for broadband detection.  Given that the exposure time is proportional to S/N$^2$, this seemingly conservative assumption will negatively impact our yield estimates when compared to a S/N $\approx5$.  However, \citet{Kasdin:2006aa} suggested that S/N $>7.1$ to ensure both a low false positive rate and low missed detection rate.  Additionally, our exposure time calculations do not include unknown systematic uncertainties, which when added in quadrature to the estimated photon noise, will reduce the real-world S/N.  We therefore do not consider our S/N$=10$ a conservative requirement.

\section{Updates to Astrophysical Assumptions}
\label{astrophysical_assumptions_section}

A number of astrophysical assumptions are required to estimate exoEarth candidate yield (see Table \ref{exoearth_params_table}).  These assumptions can have profound impacts on the yield and vary from study to study.  Specifically, values used for the typical exoEarth candidate geometric albedo vary by 50\%  and the surface brightness of a Solar System-twin disk of exozodiacal dust varies by a factor of $2.5$ \citep[e.g.,][]{2010ApJ...715..122B, 2012PASP..124..418T}.  In Appendices \ref{earth_albedo_appendix}--\ref{exozodi_appendix} we examine these assumptions, along with the surface brightness of the local zodiacal cloud.  Here we summarize those findings.

The typical albedo of an exoEarth candidate is unknown and may remain unknown prior to a direct imaging mission.  For lack of observational constraints, previous works have commonly used Earth's geometric albedo (though the value used has varied from $0.2$ to $0.3$).  We found that a Lambert phase function with geometric albedo $A_G=0.2$ is a decent approximation of Earth's reflectance, which is what we used in this work.  A geometric albedo of $A_G=0.3$ overestimates the Earth's reflectance by 50\%.  The Earth's true phase function is non-Lambertian, with a number of processes increasing the reflectance in crescent phases \citep{2010ApJ...721L..67R}.  Including these processes would lead to higher exoEarth candidate yield, but we avoided including these effects because there is no guarantee that exoEarth candidates would have atmospheres identical to Earth's.  The albedos and phase functions for exoEarth candidates may be the largest source of astrophysical uncertainty for exoEarth yield calculations.  We found yield is roughly $\propto A_G^{0.8}$ for our baseline mission assuming Lambertian spheres.

We implemented a new treatment of the local zodiacal light.  We calculated the zodiacal surface brightness for each target star as a function of ecliptic latitude.  We found that as long as observations are made at solar longitudes $\gtrsim90^{\circ}$, the conventional treatment (a uniform zodiacal surface brightness of 23 mag arcsec$^{-2}$) is adequate.  Our more realistic treatment did not significantly impact exoEarth candidate yield for the baseline mission, though it did slightly alter the prioritization of observed stars.

Finally, we investigated the assumptions that go into the calculation of exozodiacal light.  We found that complex estimates of exozodiacal light that self-consistently calculate the exozodi surface brightness with the planet's orbital orientation and phase do not significantly impact yield or target prioritization and require significant computational effort.  On the other hand, the conventional treatment of exozodiacal dust as a uniform surface brightness independent of stellar type unfairly penalizes late type stars.  We chose to define 1 ``zodi" of dust as a constant habitable zone optical depth defined at the Earth-equivalent insolation distance, such that the exozodiacal surface brightness at wavelength $\lambda$, $I_{\rm EZ}(\lambda) \propto 10^{-0.4 M_{\lambda}^{\star}}/L_{\star}$, where $M_{\lambda}^{\star}$ is the absolute stellar magnitude at wavelength $\lambda$ and $L_{\star}$ is the bolometric stellar luminosity.  Using this definition, yields are negligibly increased, but $\sim15\%$ more K and M dwarf type stars are observed.

\section{Defining a Yield Goal}
\label{yield_goals_section}

Given the caveats detailed in Section \ref{caveats}, the absolute yield numbers presented in this work will undoubtedly change as our calculations advance and our understanding of the assumptions improve.  We emphasize that the absolute yield numbers are sensitive to quantities that we do not vary in detail: poorly constrained astrophysical quantities, like $\eta_{\Earth}$ and the typical exoEarth candidate albedo, as well as assumed telescope and instrument parameters, like the required SNR for detection and the system throughput.  \emph{Rather than the absolute exoEarth candidate yield numbers, the primary focus of this paper is the relative effectiveness of the yield calculation methods, and the dependencies on varied astrophysical and mission parameters.}  Nonetheless, at some point one must decide upon a yield goal: the number of exoEarth candidates required to achieve a particular science goal.  Here we present a quantitative approach to defining the yield goal most appropriate for missions that seek to find Earth-like planets.

The parameter $\eta_{\earth}$ is often casually referred to as the fraction of stars that have Earth-like planets in their habitable zones.  Prior to planning a direct imaging mission, however, we will likely only measure the sizes of a small number of these planets.  In truth, $\eta_{\earth}$ reflects the fraction of stars that harbor Earth-sized planets in their habitable zones, an exoEarth \emph{candidate}.

Some unknown fraction of these Earth-sized planets will be barren rocks, some fraction will have liquid water on their surfaces ($\eta_{\rm H_2 O}$), and some fraction may have biomarkers consistent with the presence of life ($\eta_{\rm life}$).  Thus, if a mission's science goal is to find Earth-\emph{like} planets by searching for signs of water and/or life, the completely unconstrained quantities $\eta_{\rm H_2 O}$ and $\eta_{\rm life}$ will determine the success of the mission.  While we cannot know these quantities a priori, we can calculate how many exoEarth candidates are required to place compelling constraints on these factors.  

The binomial distribution function,
\begin{equation}
 P\left(x,n,p\right) = \frac{n!}{x! (n-x)!}\, p^x\, \left(1 - p \right)^{n-x},
\end{equation}
expresses the probability $P$ of $x$ successes out of $n$ tries, if each try has an independent probability $p$.  For the exoEarth science goal at hand, $n=N_{\rm EC}$ represents the number of exoEarth candidates, $x$ is the number of candidates that exhibit the feature for which we are searching (e.g., water), $p = \eta_x \epsilon$ is the probability of detecting the feature, $\eta_x$ is the statistical fraction of exoEarth candidates that exhibit the feature, and $\epsilon$ is our detection efficiency of the feature.

\begin{figure}[H]
\begin{center}
\includegraphics[width=4in]{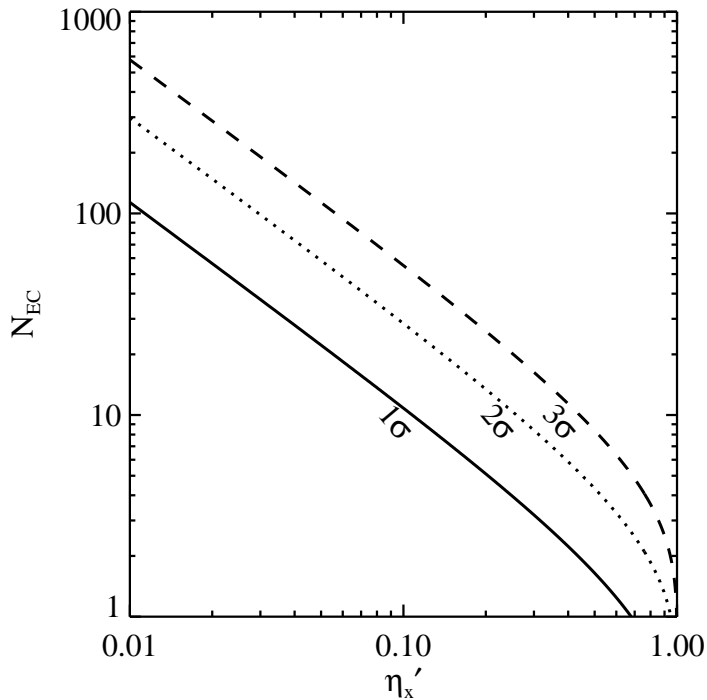}
\caption{The required number of exoEarth candidates, $N_{\rm EC}$, to constrain $\eta_x < \eta_x'$ at a given confidence level, where $\eta_x$ represents the statistical fraction of exoEarth candidates that exhibit a desired feature (e.g., water). \label{confidence_curves}}
\end{center}
\end{figure}

In the worst case scenario of a null result in which zero detections of the desired feature are made, we should require a scientifically compelling constraint on the quantity $\eta_x$.  Setting $x=0$ and assuming $\epsilon \approx 1$, we find the simple expression
\begin{equation}
 P\left(0,N_{\rm EC},\eta_x \right) = \left(1 - \eta_x \right)^{N_{\rm EC}}.
\end{equation}
Solving for $N_{\rm EC}$ gives
\begin{equation}
 N_{\rm EC} = \frac{\log{(1-C)}}{\log{(1-\eta_x)}},
\end{equation}
where $C = 1 - P$ is the confidence of the constraint on $\eta_x$.  Figure \ref{confidence_curves} shows the number of exoEarth candidates required to place 1, 2, and 3$\sigma$ constraints on $\eta_x < \eta_x'$ as a function of $\eta_x'$.  Alternatively, one could use Figure \ref{confidence_curves} to interpret the possible constraint on $\eta_x'$ given an expected number of exoEarth candidates.

We posit that a reasonably compelling limit is $\eta_x < 0.1$, whether $\eta_x = \eta_{\rm H_2 O}$ or $\eta_x = \eta_{\rm life}$.  \emph{Thus, to place a 3$\sigma$ constraint that $\eta_x < 0.1$, we require approximately 55 exoEarth candidates.}

\begin{figure}[H]
\begin{center}
\includegraphics[width=4in]{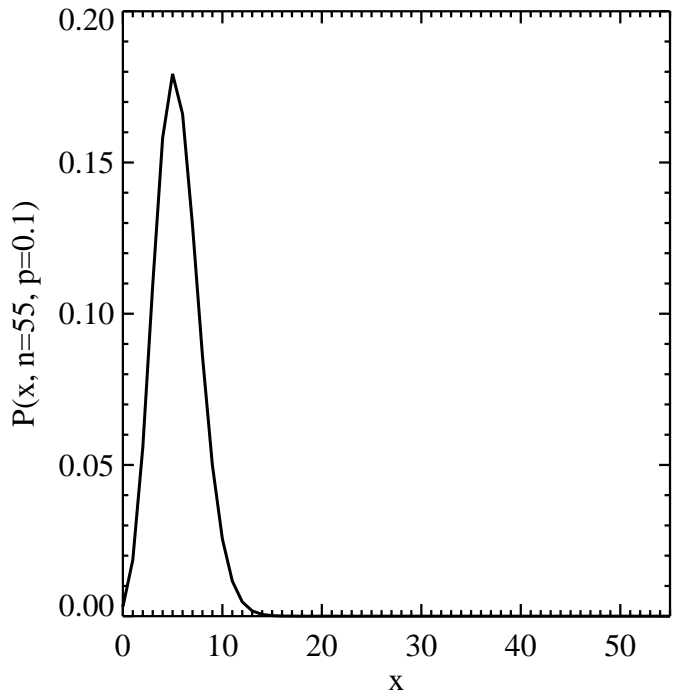}
\caption{The binomial probability distribution for $n=55$ and $p=0.1$ as a function of $x$, i.e. the probability of detecting $x$ exoEarths that exhibit a desired feature from a sample of 55 candidates assuming a statistical frequency of the feature $\eta_x = 0.1$. \label{sample_binomial}}
\end{center}
\end{figure}

Figure \ref{confidence_curves} equivalently states the number of exoEarth candidates required to detect the desired feature in at least one candidate at the given level of confidence.  \emph{Given 55 exoEarth candidates, a mission would have a $99.7\%$ chance of detecting at least one exoEarth with water if $\eta_{\rm H_2 O} \ge 0.1$ and $\epsilon \approx 1$}.  The most probable number of detections is of course equal to $\eta_x \times N_{\rm EC}$, as shown in the sample binomial distribution in Figure \ref{sample_binomial}, for which we assume $\eta_x = 0.1$ and $N_{\rm EC} = 55$.

\section{Calculating ExoEarth Candidate Yield}
\label{exoEarth_yield_section}

The fundamentals of our exoEarth candidate yield calculations are based on \citet{2005ApJ...624.1010B}.  Here we provide a brief summary of those techniques.  We divide our exoEarth candidate yield calculations into four distinct steps: target list selection, completeness calculation for each star, exposure time calculation for each star, and yield maximization.

\subsection{Target list}

To select a pool of potential targets we followed the procedures outlined in \citet{2005ApJ...624.1010B} and \citet{2012PASP..124..418T}.  We queried the \emph{Hipparcos} catalog for all stars with parallax $>20\arcsec$ ($d<50$ pc); we found the 30 pc cutoff used by \citet{2005ApJ...624.1010B} and \citet{2012PASP..124..418T} insufficient when modeling our baseline mission with apertures $\gtrsim 10$ m.  We then cross-referenced our list with the \emph{Hipparcos} Double \& Multiples Catalog and the Washington Double Star Catalog to remove stars with known companions within $10\arcsec$, leaving a total of 5449 potential targets.  Stray light from such companions may impact the required planet exposure times by increasing the background light.  For each target star, we calculated stellar luminosity from the \emph{Hipparcos} parallax data, \emph{Hipparcos} Johnson $V$ and $B-V$ values, and bolometric corrections given by Equations 1 and 2 from \citet{2012PASP..124..418T}.  As discussed in \citet{2012PASP..124..418T}, the \emph{Hipparcos} catalog is incomplete for late type stars.  Since late type stars have compact habitable zones, the \emph{Hipparcos} incompleteness may lead to underestimation of exoEarth candidate yields for large apertures with small inner working angles.

\subsection{Completeness calculator}

For exoEarth candidate yield, ``completeness" is defined as the chance of observing an exoEarth candidate around a given star if that exoEarth candidate exists \citep{2004ApJ...607.1003B}.  The chance of observing an exoEarth candidate is therefore the product of completeness and $\eta_{\earth}$.  For an exoEarth candidate to be observable, it must appear exterior to the IWA of the instrument, interior to the outer working angle (OWA), and must be bright enough to observe at a given S/N within a given exposure time.  Because we will know little about the majority of planetary systems prior to imaging them, completeness is calculated via a Monte Carlo simulation of all possible planets satisfying certain constraints.

We calculated single-visit obscurational and photometric completeness for each star in the target list using the same methods outlined in \citet{2005ApJ...624.1010B}.  We started by simulating $10^5$ exoEarths around a Sun-like star at 1 pc using the planet parameters listed in Table \ref{exoearth_params_table}.  For each synthetic exoEarth, we calculated the projected angular separation of the planet and star, $s_{\rm p}$, and the flux ratio of the planet to the star, expressed as
\begin{equation}
	\Delta {\rm mag}_{\rm p} = -2.5 \log{\frac{f_{\rm p}}{f_{\star}}},
\end{equation}
where $f_{\rm p}$ is the flux from the planet and $f_{\star}$ is the flux from the star.

\begin{deluxetable}{ccl}
\tablewidth{0pt}
\footnotesize
\tablecaption{Baseline Astrophysical Parameters\label{exoearth_params_table}}
\tablehead{
\colhead{Parameter} & \colhead{Value} & \colhead{Description} \\
}
\startdata
$\eta_{\earth}$ & $0.1$ & Fraction of stars with an exoEarth candidate \\
$R_{\rm p}$ & $1$ $R_{\earth}$ & Planet radius \\
$a$ & $[0.7,1.5]$ AU\tablenotemark{*} & Semi-major axis (uniform distribution) \\
$e$ & $[0,0.35]$ & Eccentricity (uniform distribution) \\
$\cos{i}$ & $[-1,1]$ & Cosine of inclination (uniform distribution) \\
$\omega$ & $[0,2\pi)$ & Argument of pericenter (uniform distribution) \\
$M$ & $[0,2\pi)$ & Mean anomaly (uniform distribution) \\
$\Phi$ & Lambert & Phase function \\ 
$A_G$ & $0.2$ & Geometric albedo (corresponds to spherical albedo $A_B = 0.3$) \\
$z$ & 23 mag arcsec$^{-2}$\tablenotemark{\dag}  & Surface brightness of zodiacal light \\
$x$ & 22 mag arcsec$^{-2}$\tablenotemark{\ddag}  & Surface brightness of 1 zodi of exozodiacal dust \\
$n$ & $3$ & Number of zodis for all stars \\
\enddata
\vspace{-0.1in}
\tablenotetext{*}{$a$ given for a solar twin.  The habitable zone is scaled by $\sqrt{L_{\star}/L_{\sun}}$ after calculating projected separation $s_{\rm p}$.}
\tablenotetext{\dag}{Varies with ecliptic latitude---see Section \ref{astrophysical_assumptions_section} and Appendix \ref{zodi_appendix}}
\tablenotetext{\ddag}{For Solar twin. Varies with spectral type---see Section \ref{astrophysical_assumptions_section} and Appendix \ref{exozodi_appendix}}
\end{deluxetable}

For simplicity, we require habitable zones to receive a constant bolometric stellar flux regardless of spectral type.  Thus, we only need to calculate the projected separations once; we can scale the habitable zone with stellar luminosity to maintain a constant bolometric stellar flux.  For each star on the target list, we scaled the projected separations according to
\begin{equation}
	s_{{\rm p},i} = s_{\rm p} \sqrt{\frac{L_{\star,i}}{L_{\sun}}} \left(\frac{1\, \rm{pc}}{d_i}\right),
\end{equation}
where $s_{{\rm p},i}$ is the projected separation of a single synthetic exoEarth from the $i^{\rm th}$ star, $L_{\star,i}$ is the stellar luminosity of the $i^{\rm th}$ star, $L_{\sun}$ is the solar luminosity, and $d_i$ is the distance to the $i^{\rm th}$ star.  The flux ratio of the planet and star also changes with spectral type.  For each star on the target list, we scaled the planet to star flux ratio according to
\begin{equation}
	\Delta {\rm mag}_{{\rm p},i} = \Delta {\rm mag}_{\rm p} + 2.5 \log{\frac{L_{\star,i}}{L_{\sun}}},
\end{equation}
where $\Delta \rm{mag}_{{\rm p},i}$ is the magnitude flux ratio of a single synthetic exoEarth for the $i^{\rm th}$ star.  Completeness $C_i(\Delta {\rm mag_{obs}}, \rm{IWA}, \rm{OWA})$ for the $i^{\rm th}$ star is then calculated as the fraction of synthetic exoEarths with ${\rm IWA} < s_{{\rm p},i} < {\rm OWA}$ and $\Delta {\rm mag}_{{\rm p},i} < \Delta {\rm mag_{obs}}$, where $\Delta {\rm mag_{obs}}$ expresses the faintest planet that can be detected at the desired S/N in a certain exposure time.

\subsection{Exposure time calculator}

Assuming negligible read noise and dark counts, we approximate the exposure time required to image a planet with $\Delta {\rm mag_p} = \Delta {\rm mag_{obs}}$ as
\begin{equation}
\label{tau_equation}
	\tau = \left({\rm S/N}\right)^2 \left(\frac{{\rm CR_p} + 2\, {\rm CR_b} }{{\rm CR_p}^2}\right),
\end{equation}
where $\rm{S/N}$ is the signal to noise ratio desired for the planet, ${\rm CR_p}$ is the photon count rate for the planet, ${\rm CR_b}$ is the photon count rate for the background, and $\Delta {\rm mag_{obs}} <  \Delta {\rm mag_{floor}}$, where the systematic noise floor $\Delta {\rm mag_{floor}}$ is the dimmest planet detectable at the given S/N \citep{2005ApJ...624.1010B}.  The factor of two in front of the background count rate is due to the necessity of background subtraction.  The planet count rate is given by

\begin{equation}
\label{CRp_equation}
	{\rm CR_p} = F_0\, 10^{-0.4\left({\rm m}_V + \Delta {\rm mag_{obs}}\right)}\, A\, \Upsilon\, T\, \Delta \lambda,
\end{equation}
where $F_0$ is the zero-magnitude flux at $V$ band, $m_V$ is the stellar apparent $V$-band magnitude, $A = \pi D^2 / 4$ is the effective collecting area of the telescope aperture, $D$ is the telescope diameter, $\Upsilon$ is the fraction of the diffraction limited PSF contained within the aperture defined by the angular radius $\theta = X \lambda/D$, $\lambda = \lambda_V$ is the central wavelength of the $V$ bandpass, $T$ is total facility throughput, and $\Delta \lambda$ is the bandwidth.  We assume simple aperture photometry with $X=0.7$ such that $\Upsilon=0.69$, which roughly maximizes the planet to background flux ratio.  In reality, coronagraphs will broaden the PSF of the telescope, such that the optimal $\Upsilon \ne 0.69$.  We leave a detailed analysis of how coronagraph PSF broadening will affect $\Upsilon$ to future work.  Expanding all dependencies on $\lambda$ and $D$, we find
\begin{equation}
\label{CRp_equation_expanded}
	{\rm CR_p} = F_0\, 10^{-0.4\left({\rm m}_V + \Delta {\rm mag_{obs}}\right)}\, \frac{\pi D^2}{4}\, \Upsilon\, T\, \Delta \lambda,
\end{equation}
such that the planet's count rate is proportional to the collecting area of the telescope.

We assume negligible read noise for broadband detections using future detector technologies.  Thus, three sources of background flux dominate ${\rm CR_b}$: the leaked stellar light (${\rm CR_{b,\star}}$), the local zodiacal light (${\rm CR_{b,zodi}}$), and the exozodiacal light (${\rm CR_{b,exozodi}}$).  The count rate for leaked stellar light is expressed as
\begin{equation}
\label{CRbstar_equation}
	{\rm CR_{b,\star}} = F_0\, 10^{-0.4{\rm m}_V}\, \zeta\, {\rm PSF_{\rm peak}}\, \Omega\, A\, T\, \Delta \lambda,
\end{equation}
where $\zeta$ is the uniform contrast level of suppressed starlight measured relative to the PSF peak per unit solid angle, ${\rm PSF_{\rm peak}} = \pi D^2/4 \lambda^2$ expresses the theoretical Airy pattern peak per unit solid angle under the assumption of a diffraction-limited PSF, and $\Omega = \pi(X\lambda/D)^2$ is the solid angle subtended by the photometric aperture.  Expanding all dependencies on $\lambda$ and $D$, Equation \ref{CRbstar_equation} can be expressed as
\begin{equation}
\label{CRbstar_equation_expanded}
	{\rm CR_{b,\star}} = F_0\, 10^{-0.4{\rm m}_V}\, \zeta\, \frac{\pi^2 X^2}{4}\, \frac{\pi D^2}{4}\, T\, \Delta \lambda,
\end{equation}
such that the leaked starlight's count rate is proportional to the collecting area of the telescope.

The local zodiacal light is expressed as
\begin{equation}
\label{CRbzodi_equation}
	{\rm CR_{b,zodi}} = F_0\, 10^{-0.4z}\, \Omega\, A\, T\, \Delta \lambda,
\end{equation}
where $z$ is the typical surface brightness of the zodiacal light in magnitudes per unit solid angle at $V$ band.  Expanding all dependencies on $\lambda$ and $D$, we find
\begin{equation}
\label{CRbzodi_equation_expanded}
	{\rm CR_{b,zodi}} = F_0\, 10^{-0.4z}\, \frac{\pi^2 X^2}{4}\, \lambda_V^2\, T\, \Delta \lambda,
\end{equation}
such that the local zodiacal light's count rate is independent of telescope diameter.

Finally, the exozodiacal light is given by 
\begin{equation}
\label{CRbexozodi_equation}
	{\rm CR_{b,exozodi}} = F_0\, n\, 10^{-0.4x}\, \Omega\, A\, T\, \Delta \lambda,
\end{equation}
where $x$ is the surface brightness of 1 zodi of exozodiacal light in magnitudes per unit solid angle, and $n$ is the typical number of zodis assumed for all stars.  Expanding all dependencies on $\lambda$ and $D$, we find
\begin{equation}
\label{CRbexozodi_equation_expanded}
	{\rm CR_{b,exozodi}} = F_0\, n\, 10^{-0.4x}\, \frac{\pi^2 X^2}{4}\, \lambda_V^2\, T\, \Delta \lambda,
\end{equation}
such that the exozodiacal light's count rate is independent of telescope aperture.

Note that ${\rm CR_p}$ and ${\rm CR_{b,\star}}$ are proportional to $D^2$ while ${\rm CR_{b,zodi}}$ and ${\rm CR_{b,exozodi}}$ are independent of $D$.  Thus, we may be tempted to think that exozodis will affect larger apertures less.  However, for a given star ${\rm CR_p}$ and ${\rm CR_{b,\star}}$ are proportional to $d^{-2}$ while ${\rm CR_{b,zodi}}$ and ${\rm CR_{b,exozodi}}$ are independent of stellar distance, as expected for resolved sources.  Because larger apertures have shorter exposure times, missions with larger apertures will eventually observe more distant targets, for which the exozodi problem becomes worse.  Thus the dependence of the yield on parameters like exozodi level and contrast cannot be intuited from the count rate equations above, and will be controlled by the combination of these equations and the target list selected for observation.

\begin{deluxetable}{ccl}
\tablewidth{0pt}
\footnotesize
\tablecaption{Baseline Mission Parameters\label{baseline_params_table}}
\tablehead{
\colhead{Parameter} & \colhead{Value} & \colhead{Description} \\
}
\startdata
$D$ & $8$ m & Telescope diameter \\
$\lambda$ & $0.55$ $\mu$m & Central wavelength \\
$\Delta \lambda$ & $0.11$ $\mu$m\tablenotemark{*} & Bandwidth \\
IWA & $2\lambda/D$ (28.4 mas)\tablenotemark{*} & Inner working angle \\
OWA & $15\lambda/D$ (213 mas) & Outer working angle \\
$\zeta$ & $10^{-10}$\tablenotemark{*} & Contrast level in detection region, relative to theoretical Airy peak \\
$\Delta$mag$_{\rm floor}$ & $26$ & Systematic noise floor (i.e., dimmest point source detectable at S/N)\\
$T$ & $0.2$\tablenotemark{*} & End-to-end facility throughput \\
S/N & $10$ & Signal to noise ratio required for broadband detection of a planet \\
$F_0$ & $9500$ photons cm$^{-2}$ nm$^{-1}$ s$^{-1}$ & Flux zero-point at $V$ band \\
$X$ & $0.7$ & Photometry aperture radius in $\lambda/D$ \\
$\Upsilon$ & $0.69$ & Fraction of Airy pattern contained within photometry aperture \\
$\Omega$ & $\pi(X\lambda/D)^2$ radians & Solid angle subtended by photometry aperture \\
\enddata
\tablenotetext{*}{Optimistic baseline coronagraph parameters assume future (not near-term) performance.}
\end{deluxetable}

\subsection{Yield Maximization}

The estimated exoEarth candidate yield for a mission is equal to $\eta_{\earth}$ times the total completeness obtained for all observed stars within the assumed mission lifetime.  Therefore, to calculate exoEarth candidate yield, one must decide which stars to observe, how long to observe them, and in what order, necessitating a target prioritization metric.  Completeness $C$, the chance of observing an exoEarth candidate around a given star, can be thought of as the ``benefit" of observing a star.  The ``cost" of observing a star is the exposure time $\tau$.  Thus, a reasonable prioritization metric is simply $C/\tau$, the benefit-to-cost ratio.  The calculation of this ratio appears elementary, but combining completeness with exposure time is non-trivial and the method used can significantly impact the yield estimates as well as how the yield responds to changes in mission design parameters.

\subsubsection{A Review of Prioritization Methods}

Completeness and exposure time both depend on a value of $\Delta {\rm mag}_{\rm obs}$, which defines the dimmest planet that can be detected at the desired S/N.  Traditionally, models have chosen a value of $\Delta {\rm mag}_{\rm obs}$ for each star, but the choice of $\Delta {\rm mag}_{\rm obs}$ varies widely in the literature.  \citet{2010ApJ...715..122B}  assumed a fixed value of $\Delta {\rm mag}_{\rm obs}=\Delta {\rm mag}_{\rm floor}=26$ for all stars in the target list, where $\Delta {\rm mag}_{\rm floor}$ is the systematic noise floor, beyond which a point source can never be detected at the required S/N.  We refer to this method as a strict limiting search observation (SLSO) method.  An SLSO method is both straight-forward and computationally fast.  However, exoEarth candidates around late type stars have smaller values of $\Delta {\rm mag_{p}}$, so observing late type stars to the noise floor would result in the detection of planets much smaller than 1 Earth radius; from an exoEarth candidate yield perspective, an SLSO method devotes too much observing time to late type stars.

\citet{2005ApJ...624.1010B} explored varying $\Delta {\rm mag}_{\rm obs}$ to maximize the mission yield, but kept $\Delta {\rm mag}_{\rm obs}$ the same for all stars.  We refer to this method as a tuned limiting search observation (TLSO) method.  Clearly the TLSO method performs better than the SLSO method.  However, the TLSO method still fails to take advantage of the dependence of $\Delta {\rm mag_p}$ on stellar type and, requiring many calculations of photometric completeness and exposure time, is computationally inefficient.  For typical mission parameters with modest aperture sizes, the TLSO method tunes $\Delta {\rm mag_{obs}} < 26$ such that the noise floor is never reached, but we explicitly required $\Delta {\rm mag_{obs}} \le 26$ when using the TLSO method in this paper.  

\citet{2012PASP..124..418T} ignores the contribution of the planet's Poisson noise and adjusts the exposure time for each star to roughly that required to detect an exoEarth at quadrature, effectively scaling $\Delta {\rm mag}_{\rm obs} \approx 25.2 + 2.5\log{(L_{\star}/L_{\sun})}$.  This method exploits the dependence of $\Delta {\rm mag_p}$ on stellar type, but quadrature is not necessarily the ideal phase to detect exoEarths.  We refer to this method as a stellar luminosity-adjusted (SLA) method.  We required $\Delta {\rm mag}_{\rm obs} \le 26$ for this method as well.


\subsubsection{Altruistic Yield Optimization}

The AYO method combines completeness and exposure time in a fundamentally different way from the methods discussed above.  For each star, instead of calculating just one exposure time, we calculated an exposure time for each of the $10^5$ synthetic exoEarths simulated by our completeness calculator to achieve the assumed S/N$=10$.  We then sorted the synthetic exoEarths by exposure time to obtain the completeness \emph{as a function of exposure time} for each star.  The completeness curves are then used as a metric to award exposure time to potential targets.

\begin{figure}[H]
\begin{center}
\includegraphics[width=4in]{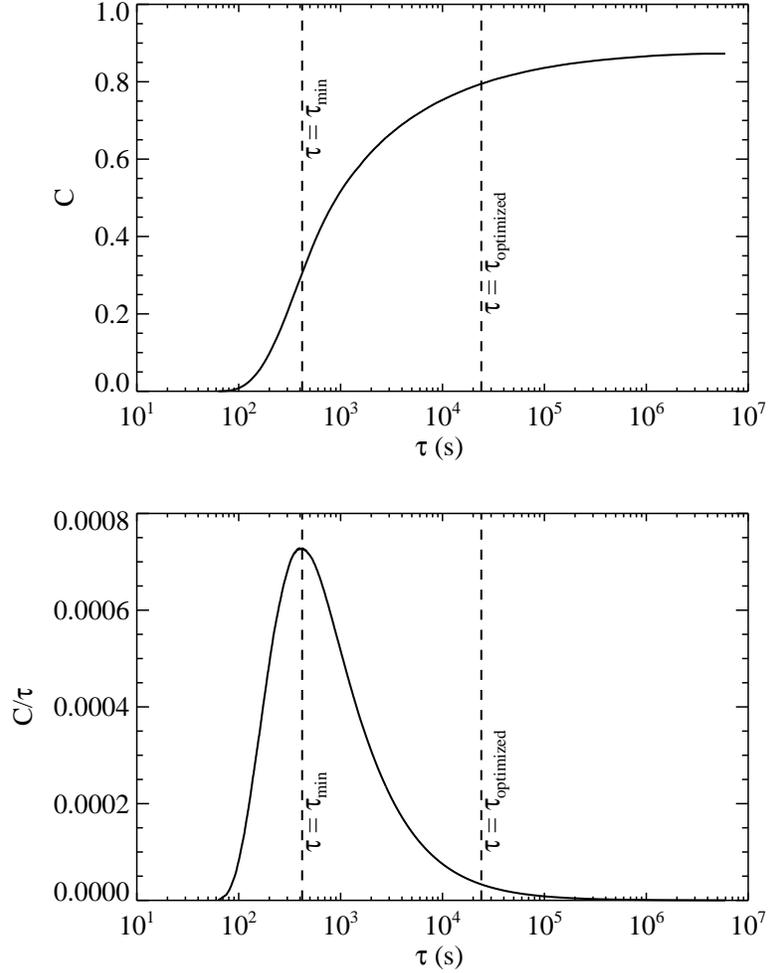}
\caption{\emph{Top}: Single-visit completeness curve for HIP 54035, calculated for our baseline mission.  \emph{Bottom}: Prioritization metric (benefit-to-cost ratio) curve.  The most efficient observation time for HIP 54035 alone occurs at $\tau = \tau_{\rm min}$.  The exposure time that helps maximize mission yield occurs at $\tau = \tau_{\rm optimized}$. \label{c_vs_t_figure}}
\end{center}
\end{figure}

The top panel in Figure \ref{c_vs_t_figure} shows an example completeness curve for HIP 54035.  This curve is similar to a vertical slice through the cumulative joint density function, as introduced by \citet{2005ApJ...624.1010B} and referred to by \citet{2013savransky}. Note that if we were to include overhead costs, they would simply be added on to the exposure time in this curve.  This curve tells us the ultimate single visit completeness for the star in the limit of $\tau \rightarrow \infty$ is $\approx0.87$.

We then divided completeness by exposure time, as shown in the bottom panel of Figure \ref{c_vs_t_figure}, to calculate the benefit to cost metric ($C_i/\tau$) as a function of $\tau$ for the $i^{\rm th}$ target.  The peak of this curve represents the most efficient exposure time for a given target, i.e., the time at which we get the most ``bang for our buck."  Observing for a shorter time would be wasteful since the efficiency and completeness are both low.  Therefore, we refer to the exposure time that maximizes efficiency for the $i^{\rm th}$ target as the minimum exposure time, $\tau_{{\rm min,}i}$, and set the exposure time for each target, $\tau_i = \tau_{{\rm min,}i}$.  We then prioritized targets by $C_i(\tau_{{\rm min,}i})/\tau_{{\rm min,}i}$ and selected the highest priority targets whose total exposure time fits within the budget of the mission.

Although we have maximized the observational efficiency of each target and prioritized the target list, we have not necessarily maximized the mission yield.  The lowest priority target selected has a long exposure time and relatively low completeness---this time might be better spent on another target.  The last step of the AYO algorithm examines the prioritized target list in reverse order and allows the lowest priority targets to ``altruistically" give their exposure time to higher priority targets.  To do so, we used the following procedure:

\begin{enumerate}
\item Remove all exposure time from the lowest priority target selected for observation.
\item Divide the exposure time into many short segments $\Delta \tau$.
\item Calculate $dC_i(\tau_i)/d\tau$ for every target.
\item Allocate $\Delta \tau$ segments to more deserving targets:
\begin{enumerate}
\item Award one $\Delta \tau$ segment to the target with the highest $dC_i(\tau_i)/d\tau$.
\item Increase $\tau_i$ for awarded target by $\Delta \tau$.
\item Recalculate $dC_i(\tau_i)/d\tau$ for awarded target.
\item Continue until all $\Delta \tau$ segments are distributed.  
\end{enumerate}
\item Repeat procedure for the next lowest priority target. 
\end{enumerate}
The algorithm continues until all targets are examined.  Figure \ref{c_vs_t_figure} shows the exposure time $\tau_{\rm optimized}$ assigned to HIP 54035 for our baseline mission after altruistic yield optimization.  A movie illustrating the AYO procedure can be downloaded from http://www.starkspace.com/code.  The exoEarth candidate detection yield is then calculated as the total completeness summed over all stars observed times $\eta_{\earth}$.

A method similar to AYO was briefly introduced by \citep{2007SPIE.6693E..22H}.  This method achieves similar results by assuming that the slopes of all completeness curves are equal at their respective optimized observing times.  While not strictly required, the assumption of equal slopes appears valid in practice due to the positive second derivative of the completeness curve prior to the inflection point.  Using this method, one must numerically solve for the ideal slope that maximizes yield, then select the prioritized target list whose cumulative exposure times fit within the mission lifetime.

\section{Results}
\label{results_section}

\subsection{AYO compared to previous methods}

\begin{figure}[H]
\begin{center}
\includegraphics[width=6in]{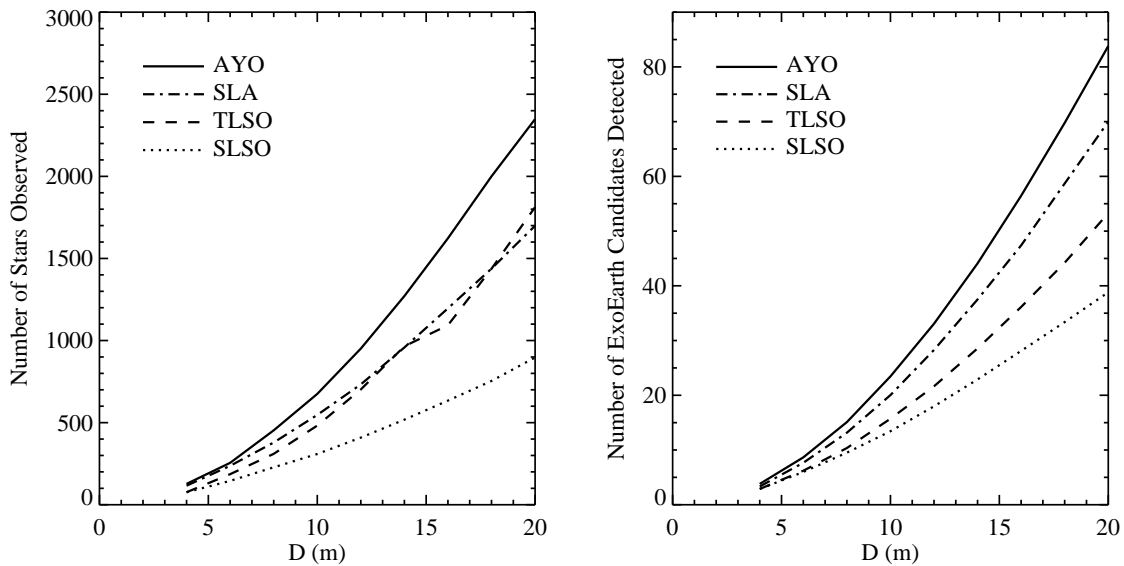}
\caption{Comparison of AYO with previous yield calculation methods for our baseline mission. \emph{Left}: Number of stars observed vs. telescope diameter.  \emph{Right}: Number of exoEarth candidates detected vs. telescope diameter.  AYO increases yield by up to $\sim$100\%.  \label{Method_Comparison-Yield_vs_D_fig}}
\end{center}
\end{figure}

The AYO method can significantly increase exoEarth candidate yield.  Figure \ref{Method_Comparison-Yield_vs_D_fig} shows the number of stars observed and exoEarth candidates detected as functions of telescope diameter for each of the methods discussed above.  The AYO method increases exoEarth candidate yield by $\sim20\%$ compared to the SLA method and $\sim100\%$ compared to the SLSO method.  The AYO method typically observes a larger number of stars for shorter periods of time, skimming only the portion of $(\Delta {\rm mag_p}, s)$ space that has the highest probability of hosting an exoEarth.  The exoEarth candidate yields for each method also respond differently to $D$.  We fit the yield vs. $D$ using functions of the form $N_{\earth}=aD^b+c$ and found $b=1.45$ for SLSO and $b=1.80$ for AYO; AYO better takes advantage of increasing telescope diameter.

\begin{figure}[H]
\begin{center}
\includegraphics[width=6in]{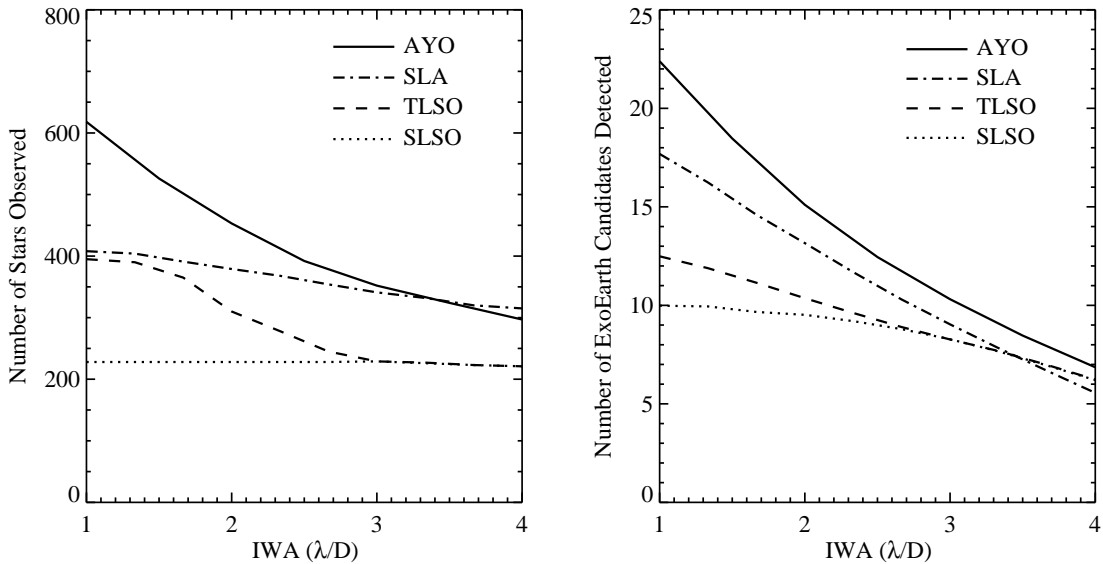}
\caption{Comparison of AYO with previous yield calculation methods for our baseline mission. \emph{Left}: Number of stars observed vs. coronagraph IWA.  \emph{Right}: Number of exoEarth candidates detected vs. coronagraph IWA.  Yields calculated with AYO respond maximally to any changes in mission parameters.  \label{Method_Comparison-Yield_vs_IWA_fig}}
\end{center}
\end{figure}

Figure \ref{Method_Comparison-Yield_vs_IWA_fig} shows the number of stars observed and exoEarth candidates detected as functions of coronagraph IWA for all methods.  The methods lead to qualitatively different IWA dependencies.  The number of stars observed is independent of IWA for the SLSO method, in contrast to the AYO method.  The SLSO method produces an exoEarth candidate yield that asymptotes at small IWA, whereas yield calculated via the AYO method continues to increase at small IWA.  Additionally, for IWA $\gtrsim 3.5 \lambda/D$, the AYO method observes fewer stars than the SLA method, but detects more exoEarth candidates.

\begin{figure}[H]
\begin{center}
\includegraphics[width=6in]{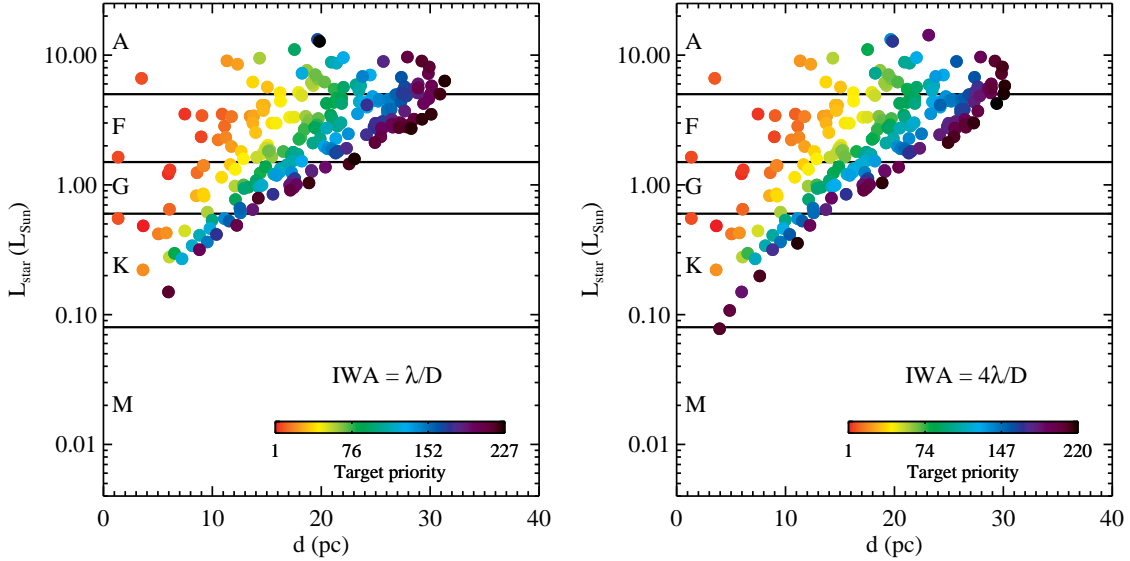}
\caption{Prioritized target list for yields calculated via the SLSO method.  \emph{Left}: ${\rm IWA}=\lambda/D$. \emph{Right}: ${\rm IWA}=4\lambda/D$.  The prioritized target list remains largely unchanged as a function of IWA. \label{Target_plot-SLSO}}
\end{center}
\end{figure}

The reason for these qualitative differences can be seen in the prioritized target lists.  Figure \ref{Target_plot-SLSO} plots the stellar luminosity as a function of distance for the SLSO prioritized target lists at ${\rm IWA}=\lambda/D$ and ${\rm IWA}=4\lambda/D$.  Each target selected for observation is plotted as a point and the color of the point corresponds to the priority rank of the target: red points are high priority and black points are low priority.  We have marked the approximate boundaries for A, F, G, K, and M main sequence stellar types for reference, though not all stars selected for observation are on the main sequence.

Although we should expect a reduced IWA to enable more observations of late type stars that have more compact habitable zones but higher planet-to-star flux ratios, there is little difference in the SLSO prioritized target lists when changing the IWA by a factor of 4.  This is because the SLSO method assumes a fixed $\Delta {\rm mag_{obs}}$ and does not take advantage of the fact that later spectral types do not require as long of exposures.  As a result, the unnecessarily long exposure time on late type stars outweighs the increase in observable habitable zone attained with a smaller IWA, preventing late type stars from moving up in priority rank.

\begin{figure}[H]
\begin{center}
\includegraphics[width=6in]{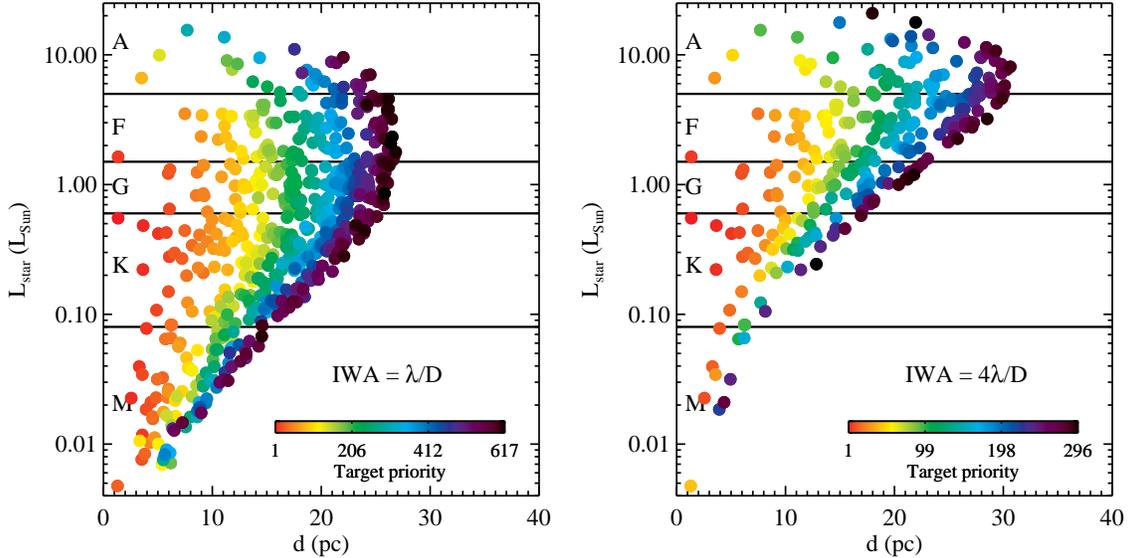}
\caption{Prioritized target list for yields calculated via the AYO method.  \emph{Left}: ${\rm IWA}=\lambda/D$. \emph{Right}: ${\rm IWA}=4\lambda/D$.  The prioritized target list changes significantly using AYO, readily adapting to changes in mission parameters. \label{Target_plot-AYO}}
\end{center}
\end{figure}

In contrast to the SLSO method, the AYO method readily adapts the prioritized target list to changes in mission parameters.  Figure \ref{Target_plot-AYO} shows the prioritized target lists for the AYO method at ${\rm IWA}=\lambda/D$ and ${\rm IWA}=4\lambda/D$.  Although the AYO method observes more stars than the SLSO method, the prioritized target list at IWA$=4\lambda/D$ looks qualitatively similar to those in Figure \ref{Target_plot-SLSO}.  However, at IWA$=\lambda/D$ the AYO prioritized target list changes dramatically, shifting preference from early type stars to late type stars, as expected given the additional access to compact habitable zones.

We note that for all prioritized target lists, the contours of constant priority resemble curves with the convex side pointing to the right, similar to a closing parenthetical sign.  The shape for these curves can be understood by moving along lines of constant stellar luminosity and distance.  Along a line of constant stellar luminosity, increasing the distance reduces the angular size of the habitable zone and increases the integration time required, explaining the fall-off in distance.  Along a line of constant distance, we see that the priority falls off both for small and large values of luminosity.  Decreasing the stellar luminosity reduces the proximity of the habitable zone to the star such that more exoEarth candidates fall within the IWA, while increasing the stellar luminosity reduces the planet-to-star flux ratio, increases the leaked stellar light, and thus the integration time.

\subsection{Response of exoEarth candidate yield to mission parameters}

We used the AYO method to study the impact of altering high-level mission parameters on exoEarth candidate yield.  To do so, we calculated the exoEarth candidate yield for $18\mathord{,}900$ sets of mission parameters.  We examined 10 values of telescope diameter $D$ equally spaced from 2 to 20 m, 9 values of coronagraph IWA equally spaced from 1--5 $\lambda/D$, 7 values of contrast $\zeta$ in the detection region from $5\times10^{-11}$ to $1.5\times10^{-9}$, and 5 values of systematic noise floor $\Delta {\rm mag_{floor}}$ ranging from $-2.5\times\log{\zeta}$ mags to $4-2.5\times\log{\zeta}$ mags.  We also examined 6 values of exozodi level $n$ ranging from 1--60 ``zodis."

We emphasize that the systematic noise floor $\Delta {\rm mag_{floor}}$ defines the \emph{dimmest planet that can be detected at the assumed S/N}.  We will sometimes express $\Delta {\rm mag_{floor}}$ as $\zeta_{\rm floor} = 10^{-0.4\Delta {\rm mag_{floor}}}$.  Because the systematic noise floor will be achieved by post-observation image processing methods, the performance of which will likely depend on $\zeta$, we will also refer to the metric $\zeta_{\rm floor}/\zeta$.

\begin{figure}[H]
\begin{center}
\includegraphics[width=6in]{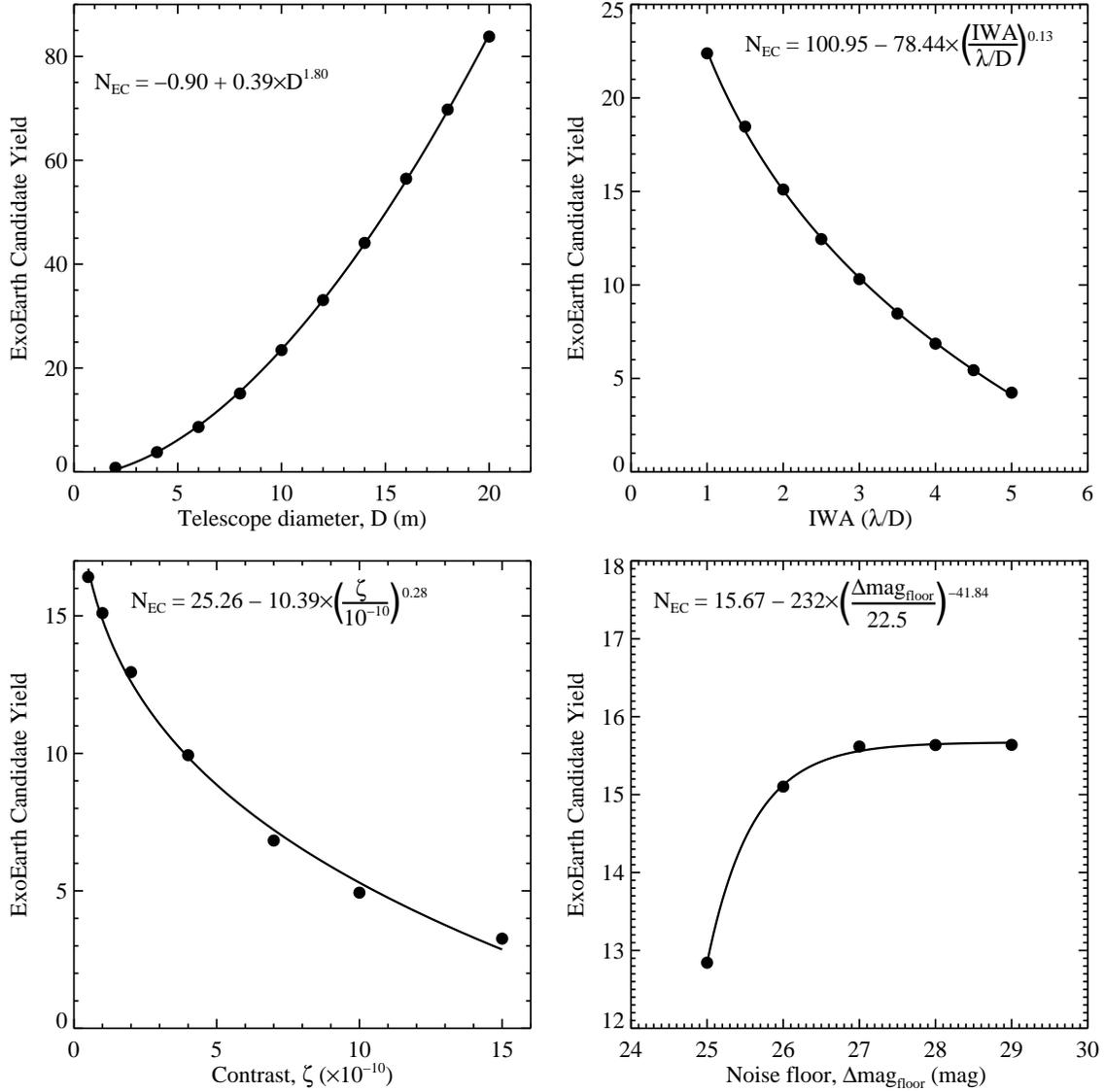}
\caption{Variations in exoEarth candidate yield from our baseline mission as we vary one telescope/instrument parameter at a time.  Calculated yields are shown as points and fits are shown as solid lines.  ExoEarth candidate yield is roughly $\propto D^{1.8}$ and plateaus at large values of systematic noise floor.  \label{AYO-Yield-multiplot_fig}}
\end{center}
\end{figure}

Figure \ref{AYO-Yield-multiplot_fig} shows the response of the exoEarth candidate yield for our baseline mission to the telescope and instrument parameters, varied one at a time, along with fits to each set of parameters.  The fits are valid only for the data points shown and do not describe the yield function over other parameter ranges.  In general, simple power laws of the form $y=ax^b$ fit the calculated yield poorly; we found that functions of the form $y=ax^b+c$ fit all yield curves well.

As shown in the upper left panel, yield depends on $D^{1.8}$ for the baseline mission.  The upper-right and lower-left plots show that yield is a function of coronagraph IWA$^{0.13}$ and $\zeta^{0.28}$.  As shown in the lower-right plot, the exoEarth candidate yield also responds strongly to the systematic noise floor for $\Delta {\rm mag_{floor}} < 26$, but plateaus for $\Delta {\rm mag_{floor}} > 26$.

Because of the constant, $c$, in our fitting functions, the magnitude of the exponent does not entirely express the sensitivity of the yield to a given parameter.  For example, doubling the IWA from our baseline mission reduces the yield by a factor of $2$ while doubling the contrast reduces the yield by a factor of $1.2$.  To judge the sensitivity of the yield to a change in a given mission parameter, we calculated the sensitivity coefficients
\begin{equation}
	\phi_x = \frac{\Delta N_{\rm EC}}{\Delta x} \frac{x}{N_{\rm EC}},
\end{equation}
where $N_{\rm EC}$ is the exoEarth candidate yield as a function of parameter $x$.  This is equivalent to the percent change in yield per percent change in parameter $x$, and in the limit $\Delta x \rightarrow 0$ is equal to the partial derivative in normalized coordinates.  We estimated $\phi_x$ at $(x, N_{\rm EC})$ using the midpoint method; given the sparseness of our grid, our values should be considered approximations.  The magnitude of $\phi$ denotes the degree of sensitivity and the sign simply denotes the directionality of the dependence.  At the baseline mission parameters, we find $\phi_D \approx 2$, $\phi_{\rm IWA} \approx -0.8$, $\phi_{\zeta} \approx -0.15$, and $\phi_{\zeta_{\rm floor}/\zeta} \approx -0.1$; for parameters near the baseline mission, the exoEarth candidate yield is most sensitive to changes in telescope diameter, followed by IWA, then contrast, and finally systematic noise floor.

Table \ref{yield_sensitivities_table} lists the approximate sensitivity coefficients for each parameter and the single-visit exoEarth candidate yield, $N_{\rm EC}$, for a sparse sampling of parameter space.  For any combination of parameters listed in Table \ref{yield_sensitivities_table}, under the assumption that the yield can be expressed as a power law function of parameter $x$ near that combination of parameters, the yield approximately scales as $N_{\rm EC} \propto x^{\phi_x}$.  The sensitivity coefficients are approximately equal to the local power law exponent.  

\begin{figure}[H]
\begin{center}
\includegraphics[width=4in]{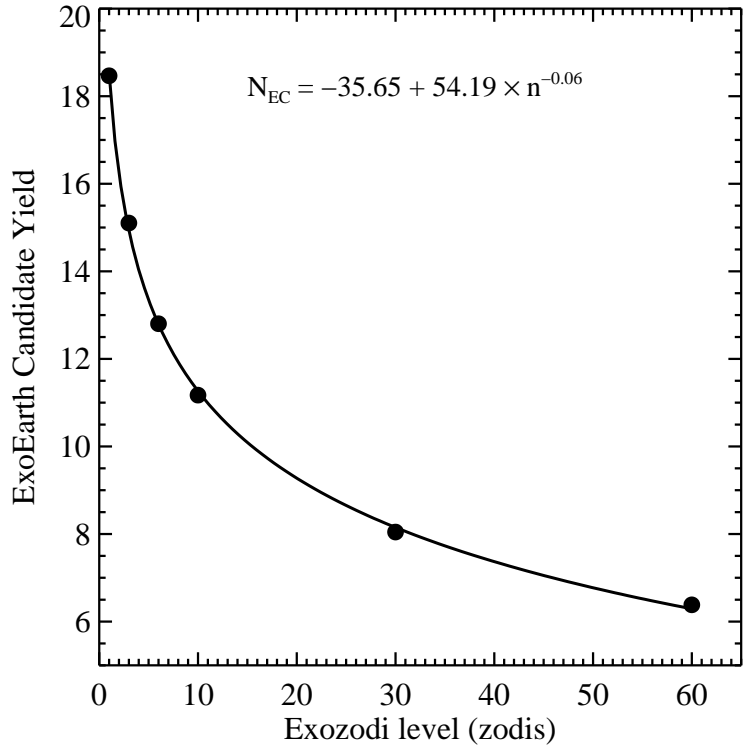}
\caption{ExoEarth candidate yield for our baseline mission as a function of exozodi level.  A factor of 10 increase in the exozodi level reduces yield by a factor of $\sim2$.\label{AYO-Yield-exozodi_fig}}
\end{center}
\end{figure}

Figure \ref{AYO-Yield-exozodi_fig} shows the exoEarth candidate yield as a function of our one varied astrophysical parameter, the level of exozodiacal dust.  A factor of 10 increase in the exozodi level reduces the yield by a factor of $\sim2$.  Thus, constraining the frequency of 10 zodi dust disks, as expected from LBTI \citep{2012PASP..124..799R}, should reduce the yield uncertainty associated with exozodi level to a factor of $\sim 2$.

\begin{figure}[H]
\begin{center}
\includegraphics[width=4in]{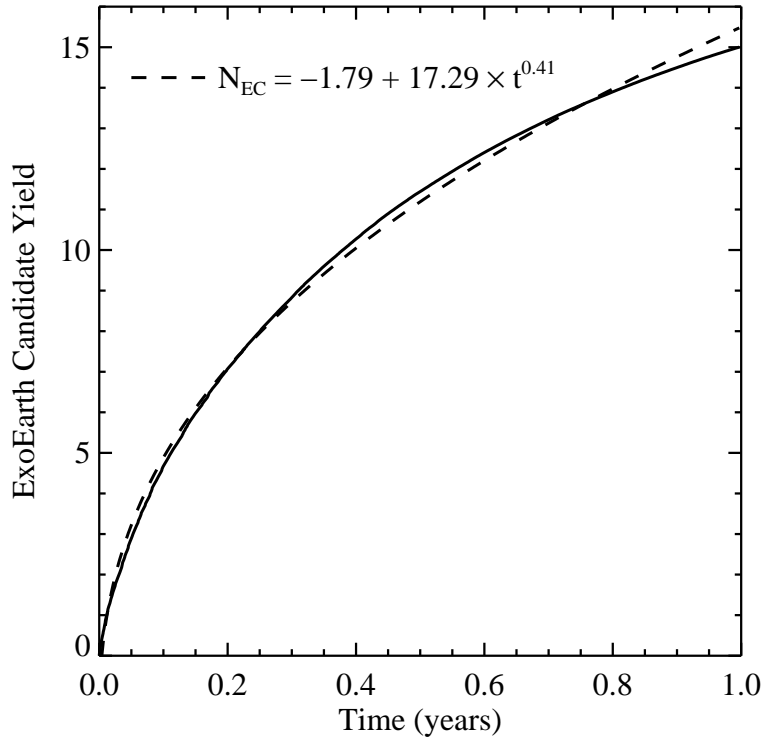}
\caption{ExoEarth candidate yield for our baseline mission over the course of the one year of total exposure time assuming all stars are observed in priority order.  The dashed line shows the best power law fit.  The yield becomes a weak function of time near the end of the mission lifetime.\label{AYO-Yield-time}}
\end{center}
\end{figure}

In Figure \ref{AYO-Yield-time}, we plot the exoEarth candidate yield as a function of time for the baseline mission, assuming that stars are observed in priority rank order over the course of the one year of total exposure time.  As the mission progresses, lower and lower priority targets are observed, such that the mission yield becomes a weaker function of time.  For $t > 0.8$ years, the exoEarth candidate yield is roughly $\propto t^{0.35}$.  Thus, modest changes to the total exposure time will not significantly impact the yield.

\begin{figure}[H]
\begin{center}
\includegraphics[width=6in]{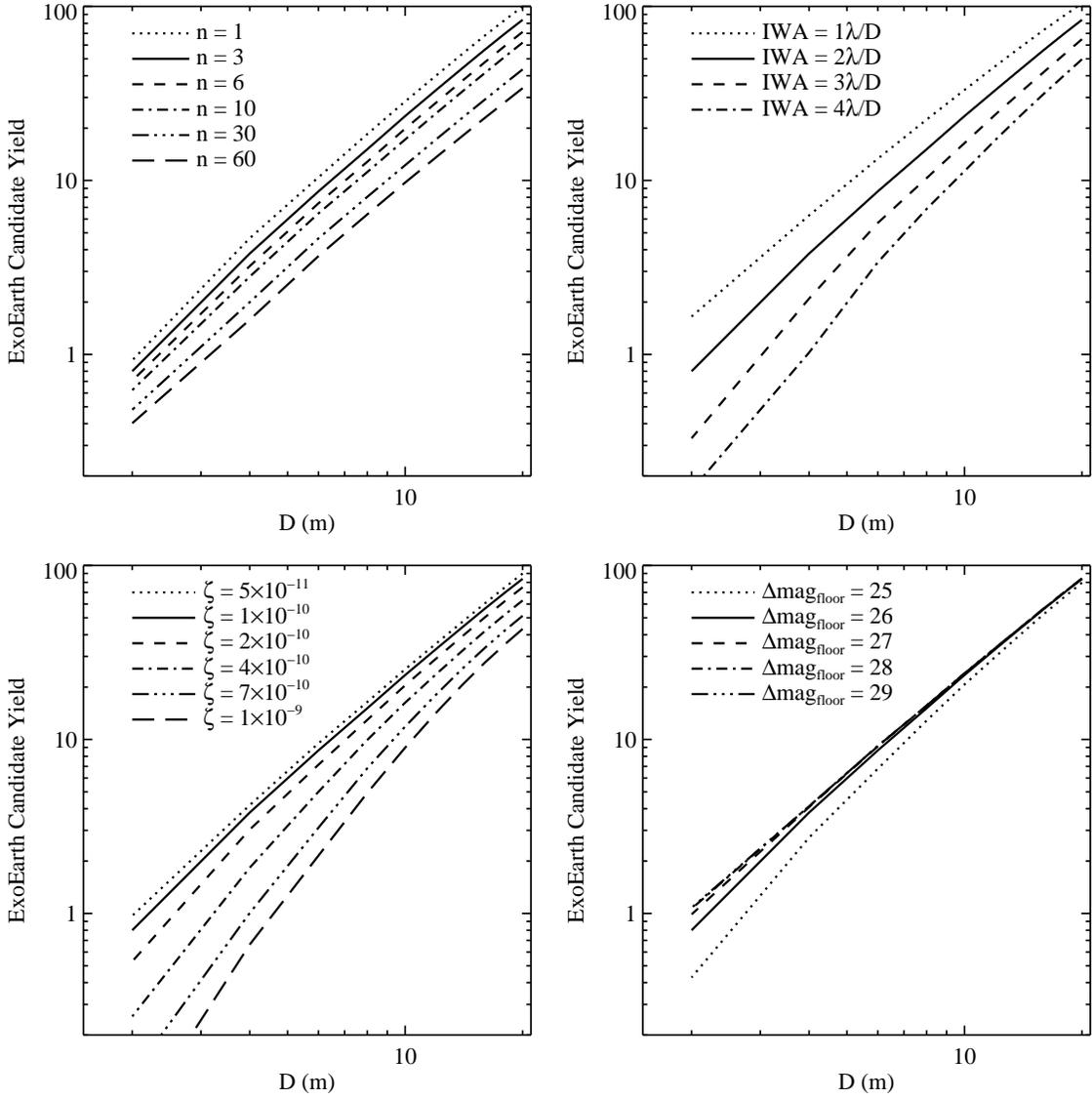}
\caption{ExoEarth candidate yield for our baseline mission as a function of $D$ while simultaneously varying exozodi level ($n$, in zodis), inner working angle (IWA), contrast ($\zeta$), and noise floor ($\Delta{\rm mag}_{\rm floor}$).  \label{AYO-Yield_vs_D-multiplot-loglog_fig}}
\end{center}
\end{figure}

As demonstrated above, the most important mission design parameter for a future direct imaging mission is the telescope diameter.  Thus, it is useful to examine how the exoEarth candidate yield responds to telescope diameter more fully.  Figure \ref{AYO-Yield_vs_D-multiplot-loglog_fig} shows plots of the exoEarth candidate yield as a function of telescope diameter while simultaneously varying one additional parameter.

The upper-left plot shows the yield as a function of telescope diameter and exozodi level.  All yield curves are parallel in log-log space, so varying the exozodi level $n$ does not impact how the yield responds to $D$.  Additionally, keeping all other parameters fixed, the relative impact of increasing the exozodi level on the exoEarth candidate yield is independent of telescope diameter; a factor of 10 increase in exozodi level reduces exoEarth candidate yield by a factor of $\sim2$, regardless of telescope diameter for our baseline mission parameters.

The remaining plots in Figure \ref{AYO-Yield_vs_D-multiplot-loglog_fig} show that the curves are not parallel in log-log space, i.e. the response of exoEarth candidate yield to $D$ changes with these parameters.  The yield is more sensitive to coronagraph IWA at smaller telescope diameters because the physical inner working angle ($\propto \lambda/D$) naturally increases for small $D$.  Because the physical IWA increases for smaller $D$, small apertures are unable to observe the brightest gibbous phase planets and are stuck observing fainter, more distant planets near quadrature, hence contrast is also more important at small $D$.  The lower-right plot shows that the plateau in exoEarth candidate yield at $\Delta {\rm mag_{floor}} \approx 26$ is roughly independent of $D$.

\begin{figure}[H]
\begin{center}
\includegraphics[width=4in]{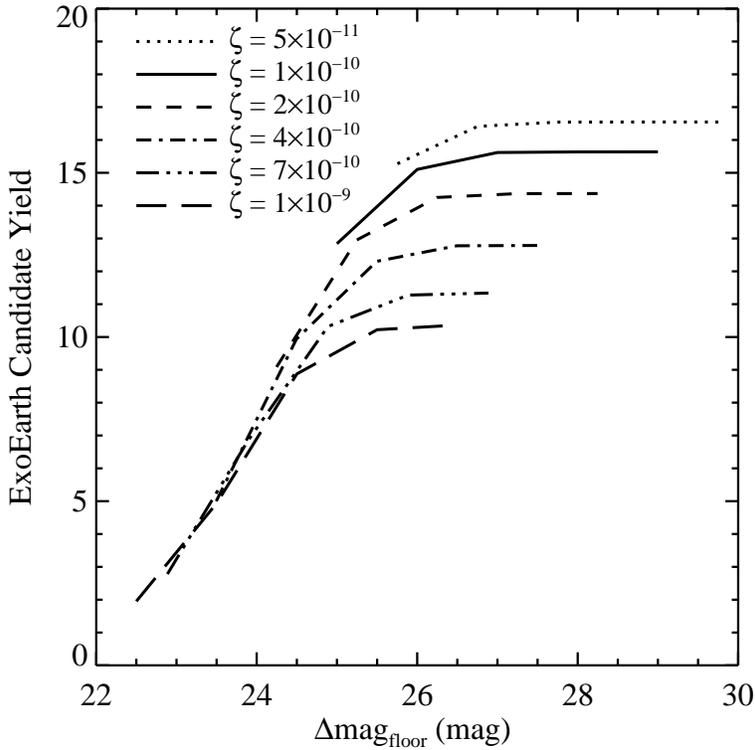}
\caption{ExoEarth candidate yield for our baseline mission as a function of systematic noise floor and contrast.  For contrasts $\sim10^{-9}$, systematic noise floors $\le 3$ magnitudes greater than the raw contrast substantially increase yield.  For contrasts $\sim10^{-10}$, systematic noise floors beyond $\Delta{\rm mag}_{\rm floor} \approx 26$ appear uneccessary.  \label{AYO-Yield_vs_limmag_vs_contrast_fig}}
\end{center}
\end{figure}

Figure \ref{AYO-Yield_vs_limmag_vs_contrast_fig} shows this asymptotic behavior of the yield for our baseline mission as a function of systematic noise floor and contrast.  At poor contrast levels, $\zeta\sim10^{-9}$, little yield is gained by obtaining systematic noise floors $\Delta {\rm mag_{floor}} > 3-2.5\log{\zeta}$, i.e. $\zeta_{\rm floor} / \zeta = 0.06$ is sufficient.  For contrasts $\zeta\lesssim10^{-10}$, a systematic noise floor $\Delta {\rm mag_{floor}} = 26$ ($\zeta_{\rm floor} = 4\times10^{-11}$) is sufficient.

We explicitly checked whether the plateau at $\Delta {\rm mag_{floor}} \approx 26$ is true for all simulations.  To do so, we produced curves similar to those shown in Figure \ref{AYO-Yield_vs_limmag_vs_contrast_fig} for every set of mission parameters.  We then divided the yield at each point in each curve by the maximum yield for that curve.  This produced curves of the fractional exoEarth candidate yield as a function of systematic noise floor.  Approximately $90\%$ of simulations with $26 < \Delta {\rm mag_{floor}} < 26.5$ have exoEarth candidate yields within $10\%$ of their maximum achievable yield.  Given our assumptions and ranges of parameters varied, a systematic noise floor $\Delta {\rm mag_{floor}} > 26$ results in little increase in planet yield.

\section{Discussion}
\label{discussion}

We find that exoEarth candidate yield is roughly $\propto D^{1.8}$ for the baseline mission.  This is more-or-less in agreement with the analytic estimate of yield $\propto D^2$ by \citet{2007MNRAS.374.1271A}, but in contrast with the analytic estimate of yield $\propto D^3$ by \citet{2008ApJ...684.1404B}.  The difference in these analytic estimates stems from the assumption of an observation time-limited sample by \citet{2007MNRAS.374.1271A} as opposed to a volume-limited sample by \citet{2008ApJ...684.1404B}.  As shown by our calculations, one year of exposure time does not achieve a volume-limited sample.  However, the analytic estimate of \citet{2008ApJ...684.1404B} suggests that increasing the exoplanet science time budget may increase the dependence of yield on $D$, a potentially important relationship that needs to be confirmed with future calculations.

Table \ref{yield_sensitivities_table} lists the sensitivity of the yield to each varied parameter over a coarse sampling of the parameter space we investigated.  As a general rule of thumb, a priority ranking of the importance of mission parameters would be $D$, IWA, $\zeta$, and then $\Delta {\rm mag_{floor}}$, with $D$ being the most important, in agreement with \citep{2007MNRAS.374.1271A}.  Of course IWA and $D$ are connected in this work, such that an increase in $D$ produces a decrease in IWA.  If IWA were independent of $D$, as would be the case for an external starshade, their impact on yield may be more equal; small IWAs are critical to achieving high exoEarth candidate yield.  A hand full of entries in Table \ref{yield_sensitivities_table} suggest that the yield can be more sensitive to coronagraph IWA than $D$ in cases with poor contrast ($\zeta \sim 10^{-9}$) and poor IWA (IWA$\approx4$).  These are likely numerical artifacts due to the small numbers of exoEarths detectable out of those simulated.

The sensitivity of the yield with respect to exozodi level is surprisingly weak.  In the limit of large $n$ (i.e., exozodi dominates the count rates) Equations \ref{tau_equation} and \ref{CRbexozodi_equation} show that the exposure time is linearly proportional to $n$ for a fixed $\Delta {\rm mag_{obs}}$.  As a result, the number of observed targets decreases.  We might naively think that the yield should therefore be proportional to $1/\tau$, or $n^{-1}$.  However, two effects work together to produce a dependence on $n$ that is substantially weaker than $n^{-1}$.

\begin{figure}[H]
\begin{center}
\includegraphics[width=6in]{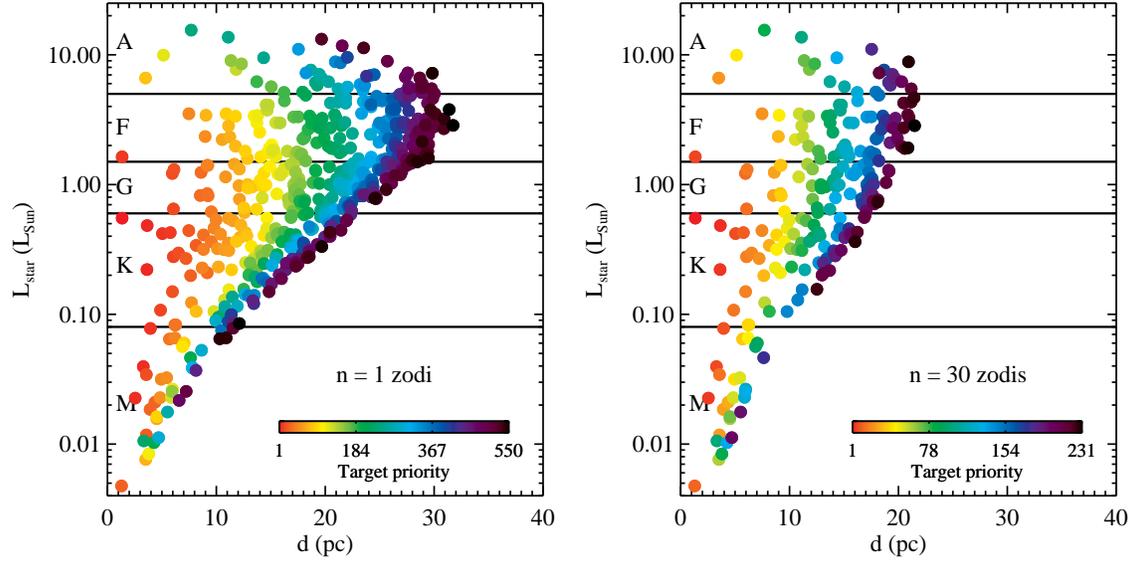}
\caption{Prioritized target list for our baseline mission as a function of exozodi level. \emph{Left:} 1 zodi of dust around every star. \emph{Right:} 30 zodis of dust around every star.  Only the nearest stars are observed in the 30 zodi case.  \label{Target_plot-AYO-exozodis}}
\end{center}
\end{figure}

First, increasing $n$ and reducing the number of observed targets preferentially eliminates poor targets.  These targets are typically distant and have longer integration times, lower completeness, and a greater exozodi count rate relative to the planet count rate.  Figure \ref{Target_plot-AYO-exozodis} shows the prioritized target lists for our baseline mission with $n=3$ and $n=30$ zodis.  Larger values of $n$ primarily eliminate distant targets that have lower planet-to-exozodi count ratios, longer exposure times, and lower completeness per star.  This selection procedure naturally biases the yield toward a weak exozodi dependence.

\begin{figure}[H]
\begin{center}
\includegraphics[width=6.5in]{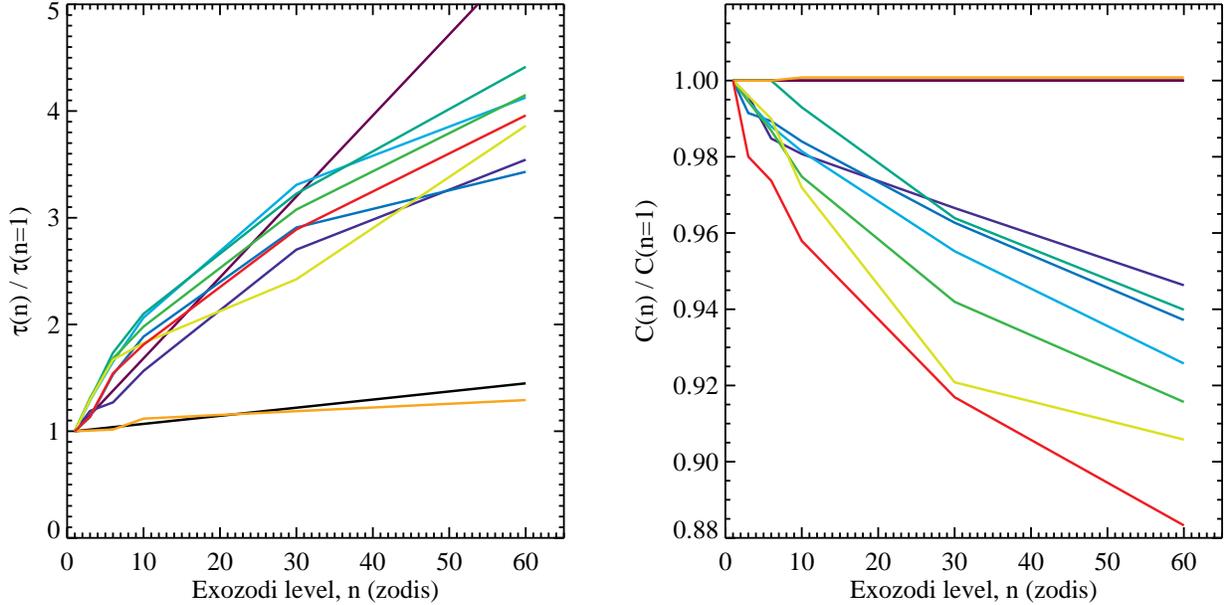}
\caption{\emph{Left:} Exposure times as a function of exozodi level, normalized to $\tau(n=1)$, for 10 high priority targets (different colored lines) using the baseline mission parameters.  \emph{Right:} Normalized completeness for the same 10 targets.  AYO adapts $\Delta {\rm mag_{obs}}$ to each star, sacrificing photometric completeness to maximize mission yield, such that the dependence of $\tau$ on $n$ is weaker than linear in some cases.\label{sample_exposure_time_vs_exozodis_figure}}
\end{center}
\end{figure}

Second, as $n$ increases, the exposure time required to achieve a given $\Delta {\rm mag_{obs}}$ increases.  For a fixed $\Delta {\rm mag_{obs}}$ for a given star, as is the case with most previous DRM codes, the calculated exposure time for that star is therefore linearly related to $n$.  However, AYO actively changes $\Delta {\rm mag_{obs}}$ for each star as the mission parameters change.  As a result, when $n$ increases, AYO can reduce $\Delta {\rm mag_{obs}}$ for a given star, sacrificing one star's photometric completeness for the benefit of the overall yield.  The left panel of Figure \ref{sample_exposure_time_vs_exozodis_figure} shows the exposure times as functions of $n$, normalized to the exposure time at $n=1$ zodi, for the 10 highest priority targets.  The right panel shows the corresponding normalized completeness curves for each star.  The completeness of some targets remains unchanged, leading to a linear relationship between $\tau$ and $n$.  AYO reduces the completeness of other targets as $n$ increases, such that $\tau$ more weakly depends on $n$ for those targets.

For our baseline mission, an increase in the exozodi level from 3 to 30 zodis reduces the yield by a factor of just $1.8$, and a change to 100 zodis would reduce the yield by a factor of just $2.8$.  In the event that LBTI constrains the frequency of exozodiacal dust disks down to the $\sim10$ zodi level, the astrophysical uncertainty in the yield associated with exozodis would be reduced to a factor of $\sim2$.  The impact of exozodi level on yield becomes even less significant for poorer contrast: increasing the exozodi level from 3 to 100 zodis for our baseline mission with a contrast of $\zeta=10^{-9}$ reduces the yield from $4.9$ to $4.0$ exoEarth candidates, a factor of 1.2 reduction.

\begin{figure}[H]
\begin{center}
\includegraphics[width=4in]{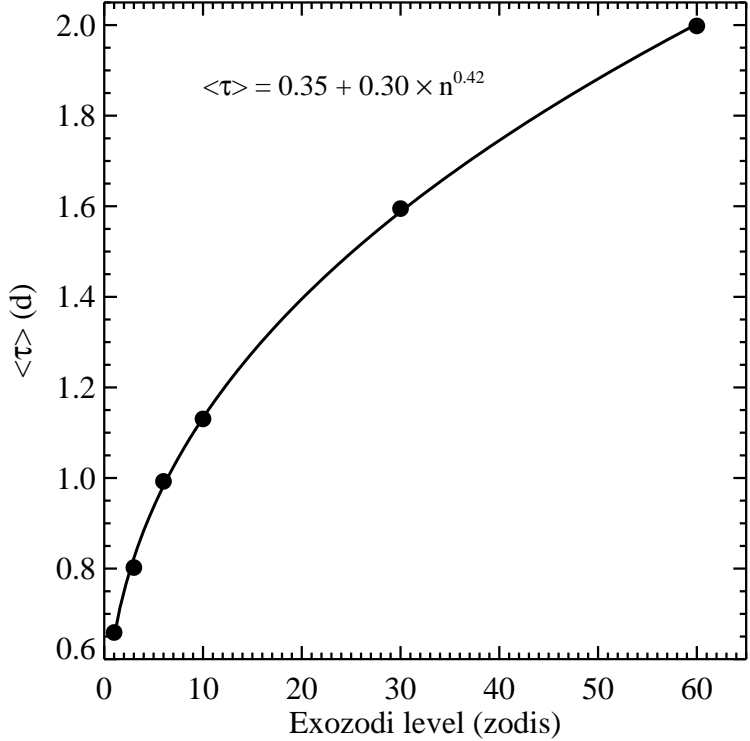}
\caption{Completeness-weighted average exposure time (optimized by AYO) as a function of exozodi level for our baseline mission.  The average optimized exposure time is a weak function of exozodi level.  \label{exposure_time_vs_exozodis_figure}}
\end{center}
\end{figure}

As expected from the optimized exposure times shown in Figure \ref{sample_exposure_time_vs_exozodis_figure} for 10 individual stars, the \emph{average} optimized exposure time is also weakly dependent on exozodi level for our baseline mission, as illustrated in Figure  \ref{exposure_time_vs_exozodis_figure}.  For example, increasing the exozodi level by a factor of 10 from 3 zodis to 30 zodis increases the average exposure time by a factor of $\sim2$.  Table \ref{yield_sensitivities_table} lists the average optimized exposure time at a variety of locations of parameter space, further illustrating the weak dependence on exozodi level.  As long as the telescope diameter $\gtrsim 8$ m and the coronagraph does not perform poorly in all parameters (i.e., IWA $<4\lambda/D$, $\zeta<10^{-9}$, $\zeta_{\rm floor}/\zeta < 0.4$), the average exposure time for detection is $\lesssim 2$ days.

Given that this work did not include revisits, we did not place a cap on the individual exposure times.  As a result, some of the exposure times, as optimized by the code, are unrealistically long.  Specifically, cases in which the coronagraph has a large inner working angle (IWA $=4\lambda/D$) and poor contrast ($\zeta=10^{-9}$) result in exposure times on the order of several tens of days or longer for some specific targets.  In these cases a large fraction of the mission lifetime is devoted to just a handful of targets, a situation that is far from ideal.  The dominant limiting factor is the coronagraph IWA; improving the obscurational completeness greatly reduces exposure times and increases the number of targets observed.

In contrast to the weak exozodi dependence, the yield appears to strongly depend on exoEarth candidate albedo, a value that may remain unconstrained prior to a direct imaging mission.  Although not varied simultaneously with other mission parameters, we found that the exoEarth candidate yield for our baseline mission was roughly proportional to $A_G^{0.8}$, such that the sensitivity coefficient $\phi_{A_G} = 0.78$ at $A_G=0.2$, roughly 3 times the magnitude of $\phi_n$.  \emph{The uncertainty in the yield associated with the typical exoEarth candidate albedo could be significantly larger than the uncertainty associated with typical exozodi level}.

It is possible that this weak exozodi dependence is related to several simplifications we have made in our calculations, specifically the lack of revisits, spectral characterization time, and overheads, and we must verify this relationship in future calculations.  Overheads may increase the impact of the exozodi on yield.  Currently there is no penalty for the number of stars observed, and AYO prefers to observe many stars for shorter exposure times to skim the most easily observable phase space; loosely speaking, AYO favors obscurational completeness over photometric completeness.  Once overheads are budgeted, AYO may be forced to observe fewer stars for longer times, thus increasing the impact of the exozodi.  On the other hand, we expect that including revisits may further reduce the exozodi dependence, since revisits can trade off photometric completeness for obscurational completeness.


Our simulations suggest that systematic noise floors $\zeta_{\rm floor} < 4\times10^{-11}$ are unnecessary for contrasts $\lesssim 10^{-10}$.  For contrasts $\sim 10^{-9}$, $\zeta_{\rm floor}$ may not need to be less than $\sim0.06\zeta$.  We again remind the reader that our definition of systematic noise floor $\Delta{\rm mag_{floor}}$ is the dimmest planet that can be detected at an S/N $=10$, and we define $\zeta_{\rm floor} = 10^{-0.4\Delta{\rm mag_{floor}}}$.

In spite of their more compact habitable zones, the success rate of finding exoEarth candidates around K and M type stars is 37\% for the baseline mission, compared to the 29\% success rate for more massive stars.  This does not incoporate the fact that exoEarth candidates may be more common around M stars, with estimates converging on $\eta_{\earth} \approx 0.4$ for M stars \citep{2013A&A...549A.109B,2013ApJ...767L...8K}.  Because of their compact habitable zones, only nearby M stars are selected for observation such that their exposure times are relatively short, $\sim 9$ hours for the baseline mission.  In addition, their planet to star flux ratios are more favorable.  Using our definition of 1 zodi being a constant habitable zone optical depth, M stars also have dimmer exozodi at V band (by a factor of $\sim2.5$).  Finally, the orbital period for exoEarths around M type stars are shorter, imposing fewer constraints on the field of regard and observation timing for a future mission.

However, characterizing exoEarth candidates around very late type stars in the near-infrared will be exceedingly challenging, if at all possible.  To keep the habitable zones of M stars accessible at 2 $\mu$m, the telescope diameter must increase by a factor of 4, or the coronagraph must work 4 times closer to the star, or a combination of the two.  Given our baseline mission parameters of $D=8$ m and IWA$=2\, \lambda/D$, we find this unlikely.  Unless an additional technology enabling even smaller IWAs is considered, such as an external starshade, a space-based observatory will likely not characterize exoEarth candidates around M stars in the near-IR.

It is important to remember that these exoEarth candidate yield calculations are based on single visit completeness and do not include revisits, integration time for spectral characterization, or overhead costs.  Adding these components will change the yield numbers and may change the mission parameter scalings and sensitivities.

\section{Conclusions}
\label{conclusions}

We created a new exoplanet yield calculator that employs an Altruistic Yield Optimization (AYO) algorithm.  AYO selects the target list and exposure times to maximize exoplanet yield and adapts maximally to changes in mission parameters.  Optimization methods like AYO can increase exoEarth candidate yield by as much as 100\% compared to previous methods.

We used our code to estimate the exoEarth candidate yield for a large suite of direct imaging mission parameters based on single-visit completeness calculations, summarized in Table \ref{yield_sensitivities_table}.  We found that exoEarth candidate yield responds most strongly to telescope diameter, followed by coronagraph inner working angle, followed by coronagraph contrast, and finally contrast floor.  We suggest that the dimmest planet required for detection at an S/N $=10$ is 26 magnitudes fainter than its host star.  For today's coronagraph contrast levels $\sim10^{-9}$, the dimmest planet is slightly brighter, $25.5$ magnitudes fainter than its host star.  However, this constraint on the noise floor likely scales with the typical, unknown exoEarth candidate albedo.

The dependence of exoEarth candidate yield and optimized exposure time on exozodiacal dust level is surprisingly weak at all telescope diameters (a factor of 10 increase in exozodi level decreases yield by $\sim2$).  This weak exozodi dependence results from the combination of a natural target selection bias and the tendency for AYO to sacrifice photometric completeness on a given target for overall yield maximization.  The dependence of yield on exoEarth candidate albedo, a parameter that may go unconstrained prior to a direct imaging mission, is comparatively strong.  Future work that incorporates revisits, spectral characterization time, and overheads into our yield estimates are required to verify this weak exozodi dependence.

Similarly, additional work is required to better constrain the absolute exoEarth candidate yield numbers presented in this paper.  Regardless, we suggest a target yield of 55 exoEarth candidates, which statistically ensures at least one exoEarth candidate with water at the 99.7\% confidence level if the fraction of exoEarth candidates with water $\eta_{\rm H_2 O} \ge 0.1$ and the detection efficiency of water is $\approx 1$.

\acknowledgments

The authors are thankful for enlightening conversations with Karl Stapelfeldt, Michael McElwain, Ruslan Belikov, and Stuart Shaklan, as well as the anonymous referee whose suggestions substantially improved the clarity of this manuscript.  This research was supported by appointments to the NASA Postdoctoral Program at Goddard Space Flight Center and Ames Research Center, administered by Oak Ridge Associated Universities through a contract with NASA.  A.R. and A.M. acknowledge support by GSFC's internal research and development fund.

\begin{appendices}

\section{ExoEarth Candidate Albedo and Phase Function}
\label{earth_albedo_appendix}

For lack of better information, exoEarth candidate yield calculations regularly use Earth's albedo and phase function for all synthetic exoEarth candidates.  It's unclear as to whether this is a reasonable assumption.  Estimates of $\eta_{\earth}$ typically include all planets in the habitable zone ranging from $\sim0.5$--$1.4$ R$_{\earth}$, which likely have a wide range of albedos and are not necessarily Earth-like.  In the Solar System, the approximate Bond albedos of Mars, Venus, and Mercury are $0.21$, $0.76$, and $0.12$, respectively \citep[e.g.][]{1977JGR....82.4249K,1985AdSpR...5..197M,1988merc.book...37V}.  We have almost no constraint on the typical albedo of an Earth-sized planet, and may have no additional observational constraints until such planets are directly imaged.  In light of this, we continue to use the na\"{\i}ve assumption that the typical exoEarth candidate albedo and phase function is Earth-like.

The Bond albedo, defined as the bolometric spherical albedo, of Earth is widely reported as $A_B = 0.3$ \citep[e.g.,][]{2001GeoRL..28.1671G}, with a similar V band spherical albedo $A_S$.  For a Lambertian sphere, the geometric albedo $A_G = (2/3)\times A_S$; if the Earth's phase function is Lambertian, then the appropriate V-band geometric albedo is $A_G=0.2$.  But is the Earth's phase function Lambertian?

\begin{figure}[H]
\begin{center}
\includegraphics[width=4in]{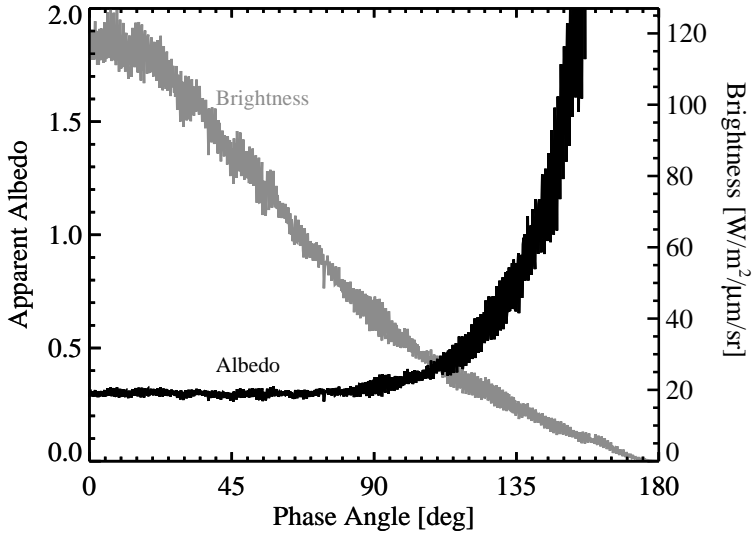}
\caption{Observation-validated model of Earth's apparent albedo (black) and brightness (gray) at V-band as a function of phase angle \citep{2010ApJ...721L..67R}.  At phase angles $<90^{\circ}$, where most exoEarth candidates would be detected, the apparent albedo is flat; a Lambert sphere with spherical albedo of $0.3$ and geometric albedo of $0.2$ is a good fit to Earth's brightness curve.  \label{earth_albedo_fig}}
\end{center}
\end{figure}

Figure \ref{earth_albedo_fig} shows the modeled V-band apparent albedo for Earth as a function of phase angle along with a modeled brightness for Earth.  Calculations of Earth's phase-dependent apparent albedo come from the Virtual Planetary Laboratory's three-dimensional, line-by-line, multiple-scattering spectral Earth model \citep{2011AsBio..11..393R}.  This model incorporates realistic absorption and scattering by the atmosphere and surface, including specular reflectance from the ocean, and direction-dependent scattering by clouds.  Data from Earth-observing satellites are used to specify the time- and location-dependent state of the surface and atmosphere.  The model has been extensively validated against observations of our planet, including data from NASA's LCROSS and EPOXI missions \citep{2011AsBio..11..393R,2014robinson_et_al}, as well as Earthshine observations of Earth's phase-dependent reflectivity \citep{2010ApJ...721L..67R}.

The apparent albedo is the effective spherical albedo required to reproduce the brightness curve under the assumption of a Lambertian phase function.  For a Lambertian sphere, the apparent albedo would be constant and equal to the true spherical albedo.  For phase angles $<90^{\circ}$, the apparent albedo of Earth is roughly constant and equal to $0.3$.  Thus, \emph{for phase angles $<90^{\circ}$ the Earth can be approximated by a Lambertian sphere with spherical albedo $A_S = 0.3$ and geometric albedo $A_G=0.2$.}  However, for phase angles $>90^{\circ}$, direction-dependent scattering from Earth's atmosphere and surface increases the brightness in crescent phases relative to a Lambertian sphere, interpreted as a much higher apparent albedo.  Given that phase angles $<90^{\circ}$ correspond to the brightest gibbous phases, the phases at which exoEarth candidates will likely be detected, approximating exoEarth candidates as a Lambertian sphere with V-band geometric albedo $A_G=0.2$ should be a reasonable choice. 

We tested the impact of including direction-dependent scattering on exoEarth candidate yield.  Assuming a Lambertian phase function, we calculated the exoEarth candidate yield for a small sample of mission parameters using a geometric albedo of $0.2$ and compared it to the exoEarth candidate yield calculated using $(2/3)$ times the apparent albedo shown in Figure \ref{earth_albedo_fig}.  We found that including direction-dependent scattering increases exoEarth candidate yield by just $\sim10\%$, validating our approximation of Earth's phase function as Lambertian.

Models suggest that most Solar System planets should exhibit qualitatively similar non-Lambertian phase functions \citep{2005ApJ...627..520S}.  However, given the large uncertainty in typical albedo, it is difficult to justify including this phase function complexity.  Therefore, for all calculations in this work we use a Lambertian sphere with geometric albedo $A_G=0.2$.

We note that exoEarth candidate albedo and phase function together are among the largest sources of astrophysical uncertainty regarding exoEarth candidate yield calculations.  For example, under the assumption of a Lambertian phase function, increasing the geometric albedo from $A_G=0.2$ to $A_G=0.3$ increases exoEarth candidate yield by 40\%.  For our baseline mission, we find that yield is roughly $\propto A_G^{0.8}$.

\section{Zodiacal Surface Brightness}
\label{zodi_appendix}

Previous calculations of exoEarth candidate yield have typically assumed a fixed, uniform value for the zodiacal light surface brightness $z$.  This is not strictly correct, since the surface brightness of the zodiacal cloud changes with the target star's ecliptic latitude and its longitude relative to the Sun.  We therefore tried using a more realistic treatment for the zodiacal cloud surface brightness.

The V-band surface brightness of the local zodiacal cloud changes with ecliptic latitude $\beta$, longitude relative to the Sun $\Delta\lambda_{\Sun}$, and observation location \citep[e.g.][]{1974A&A....30..411L}.  Assuming observations are made from a location near Earth, the V-band surface brightness of the zodiacal light is simply a function of ecliptic latitude and solar longitude,
\begin{equation}
	I_{\rm ZL} = I\!\left(\Delta\lambda_{\sun},\beta\right),
\end{equation}
The zodiacal light reaches a minimum near $\Delta\lambda_{\sun} \approx 135^{\circ}$ and has a V-band surface brightness of $23.33^{+0.06}_{-0.05}$ mag arcsec$^{-2}$ at the ecliptic poles \citep{1998A&AS..127....1L}.  We assume that most observations can be made within a few tens of degrees of $\Delta\lambda_{\sun}\!=\!135^{\circ}$, where the zodiacal light surface brightness changes slowly with $\Delta\lambda_{\sun}$.  The zodiacal light surface brightness as a function of wavelength can then be calculated as
\begin{equation}
\label{ZL_SB_equation}
	I_{\rm ZL} = \frac{I\!\left(\Delta\lambda_{\sun}\!=\!135^{\circ},\beta\!=\!0^{\circ}\right)}{I\!\left(\Delta\lambda_{\sun}\!=\!90^{\circ},\beta\!=\!0^{\circ}\right)} I\!\left(\Delta\lambda_{\sun}\!=\!90^{\circ},\beta\!=\!0^{\circ}\right) f_{135}\!\left(\beta\right),
\end{equation}
where $f_{135}(\beta)$ expresses the variation in the zodiacal light brightness as a function of ecliptic latitude at $\Delta\lambda_{\sun}=135^{\circ}$.

Using Table 17 of \citet{1998A&AS..127....1L} we find
\begin{equation}
	\frac{I\!\left(\Delta\lambda_{\sun}\!=\!135^{\circ},\beta\!=\!0^{\circ}\right)}{I\!\left(\Delta\lambda_{\sun}\!=\!90^{\circ},\beta\!=\!0^{\circ}\right)} = 0.69,
\end{equation}
and we fit the latitudinal surface brightness variation with the third degree polynomial
\begin{equation}
	f_{135}\!\left(\beta\right) \approx 1.02 - 0.566\sin{|\beta|} - 0.884\sin^2 |\beta| + 0.853\sin^3 |\beta|.
\end{equation}
Table 19 in \citet{1998A&AS..127....1L} lists the product $I(\Delta\lambda_{\sun}\!=\!90^{\circ},\beta\!=\!0^{\circ}) f_{\rm abs}$ as a function of wavelength.  At $\lambda = 0.5$ $\mu$m, $f_{\rm abs} = 1$, such that 
\begin{equation}
\label{I90_equation}
	I(\Delta\lambda_{\sun}\!=\!90^{\circ},\beta\!=\!0^{\circ}) = 5.1\times10^{-29}\text{ erg s}^{-1}\text{ cm}^{-2}\text{ Hz}^{-1}\text{ arcsec}^{-2}.
\end{equation}

Together Equations \ref{ZL_SB_equation}--\ref{I90_equation} define our calculation of zodiacal light surface brightness for each target star as a function of ecliptic latitude $\beta$.  Using these equations, approximately valid near V-band only, we calculate a surface brightness of $23.4$ mag arcsec$^{-2}$ at the ecliptic poles and $22.5$ mag arcsec$^{-2}$ in the ecliptic plane.  These equations produce a median zodiacal surface brightness for our target list of $23.0$ mag arcsec$^{-2}$, equal to the uniform surface brightness conventionally used \citep[e.g.,][]{2005ApJ...624.1010B,2012PASP..124..418T}.

To examine the impact of our more realistic treatment of the zodiacal light, we calculated the exoEarth candidate yield using the above expressions and compared it to the conventional treatment of a uniform surface brightness of $23$ mag arcsec$^{-2}$.  We find no significant change in the exoEarth candidate yield, even at low exozodi levels where the zodiacal cloud contributes a greater fraction of the background.  Upon more detailed inspection, we do find small changes in the target prioritization, with targets at high ecliptic latitudes shifting to higher priority and targets at low ecliptic latitudes shifting to lower priority ($\sim5\%$ change in rank).  Though our more realistic treatment of zodiacal light does not appear to impact exoEarth candidate yield, it requires a trivial amount of computational effort, so we implemented it for all of our calculations.

\section{Exozodiacal Surface Brightness}
\label{exozodi_appendix}

Previous works have assumed a fixed, uniform value for the exozodiacal surface brightness $x$ for 1 zodi of dust, ranging from 23 mag arcsec$^{-2}$ \citep{2005ApJ...624.1010B} to 22 mag arcsec$^{-2}$ \citep{2012PASP..124..418T}.  However, a uniform value does not account for the $1/r^2$ illumination factor we might expect for an optically thin disk.  As a result, exoEarth candidates near the outer edge of the habitable zone are unfairly penalized while exoEarth candidates near the inner edge are unfairly favored.  In addition, earlier workers have not scaled the surface brightness with stellar type.  As a consequence, under this assumption of constant surface brightness, the optical depth of 1 zodi of dust varies from star to star.  Here we examine the impact of these approximations.

\subsection{Impact of disk geometry}

We expect exozodiacal dust to be disk-like in geometry, such that disks viewed face-on should appear dimmer than those viewed edge-on.  Assuming coplanar alignment of the planetary orbits and exozodiacal disk mid-planes, the surface brightness of the exozodiacal cloud should change with the orbital configuration of \emph{each synthetic exoEarth}.  Thus, a fixed value for surface brightness unfairly penalizes face-on orbits as well as distant exoEarth candidates due to the missing $1/r^2$ illumination factor.  

It's worthwhile to note that disks viewed face-on should appear dimmer than disks viewed edge-on for two reasons.  First, assuming the disk's radial width exceeds its scale height, the line of sight-integrated surface density of an exozodiacal cloud should be higher for edge-on disks than face-on disks, giving rise to a purely geometric dimming of face-on disks.  Second, extrasolar debris disks show signs of preferential forward scattering of starlight \citep[e.g.][]{2005Natur.435.1067K,2006ApJ...650..414S,2008ApJ...673L.191D,2014stark_et_al}, qualitatively consistent with what is expected from Mie theory.  The zodiacal cloud also exhibits preferential forward scattering of sunlight \citep{1985A&A...146...67H}.  Preferential forward scattering will make face-on disks appear even dimmer since the majority of starlight is scattered into unobserved directions.  Furthermore, for moderately inclined disks, the forward scattering portion of the disk will be in front of the plane of the sky, the region in which exoEarth candidates are in crescent phase and are dimmest.

To take these effects into account, we must assume something about the geometry of the exozodiacal disk.  For lack of better information, we assumed that exozodiacal clouds are analogous to our zodiacal cloud.  We used ZODIPIC \citep{2004ApJ...612.1163M} to estimate the V band surface brightness of the zodiacal cloud around a solar twin viewed from 1 pc as a function of inclination and projected position.

We then defined 1 zodi as an exact twin of our zodiacal cloud's surface brightness distribution, scaled in circumstellar distance by $\sqrt{L_{\star}/L_{\sun}}$ to reflect the spectral type dependency of the habitable zone.  To calculate $n$ zodis of exozodiacal light, we multiplied the surface brightness of our 1 zodi model by $n$.  We note that our definition of 1 zodi here is still based upon V band surface brightness, not surface density.  As a consequence, the surface density associated with ``1 zodi" of exozodiacal light changes with spectral type, just like in the case of using a constant surface brightness.

We calculated 180 models of the zodiacal cloud to cover the full range of inclinations and saved the 2D surface brightness map of each model.  We then calculated the appropriate exozodiacal surface brightness for each synthetic exoEarth by indexing the zodiacal cloud map with the planet's inclination and 2D projected coordinates.  For each star, we adjusted the scale of the zodiacal cloud map to reflect changes in the habitable zone location prior to indexing.

\begin{figure}[H]
\begin{center}
\includegraphics[width=6.5in]{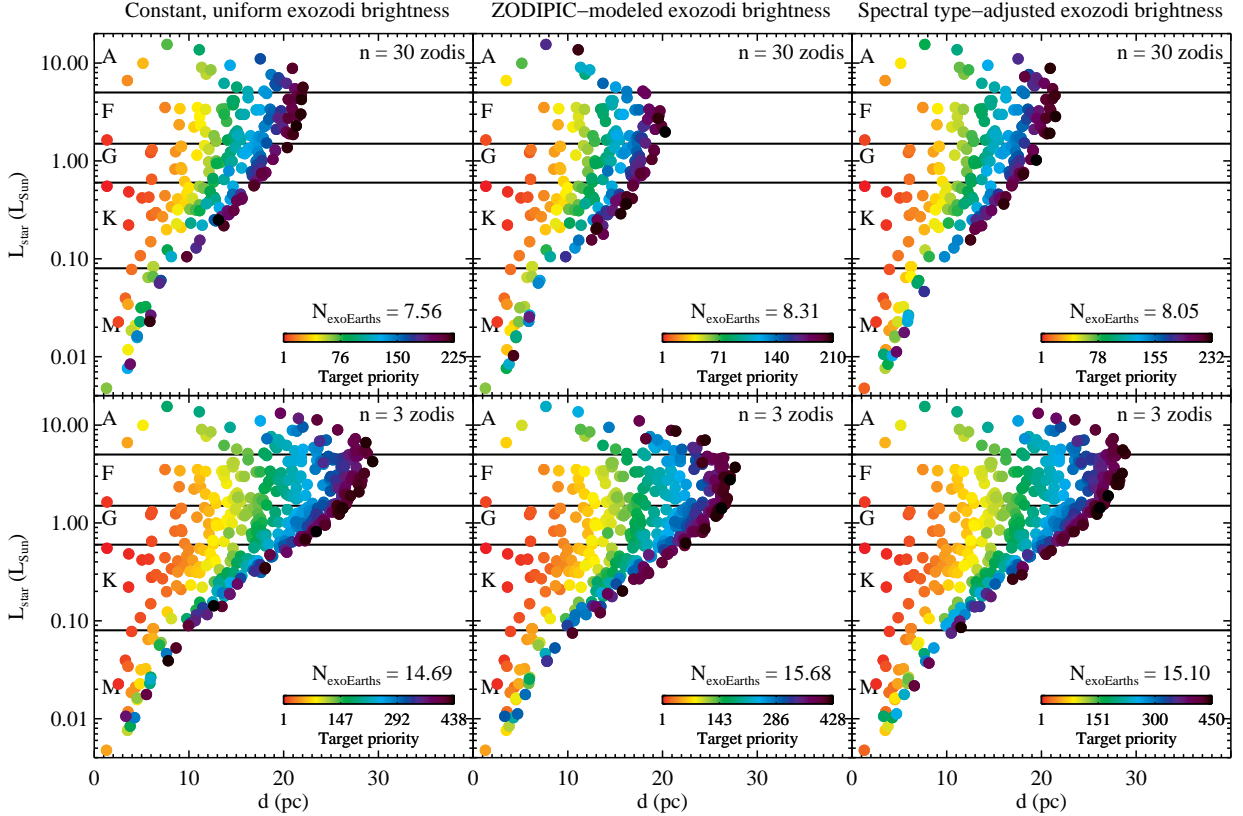}
\caption{Prioritized target lists for the baseline mission.  The top row corresponds to $n=30$ zodis of dust and the bottom row corresponds to $n=3$ zodis of dust.  \emph{Left column:} 1 zodi defined as a constant surface brightness $x=22$ mag arcsec$^{-2}$.  \emph{Middle column:} 1 zodi calculated self-consistently with planet orientation and location via ZODIPIC.  \emph{Right column:} 1 zodi defined as a constant optical depth at the EEID, the definition we chose.  Details of the exozodi treatment only affect planet yields at the few percent level and minimally alter the prioritized target list, most notably raising the priority of later type stars for the spectral type-adjusted approach.\label{exozodi_treatment_comparison_fig}}
\end{center}
\end{figure}

Figure \ref{exozodi_treatment_comparison_fig} compares the new ZODIPIC method of exozodi surface brightness calculation with previous methods.  The left panels show the prioritized target lists for our baseline mission assuming the constant $x=22$ mag arcsec$^{-2}$ value used by \citet{2012PASP..124..418T} for $n=3$ zodis of exozodiacal dust and $n=30$ ``zodis."  The middle panels show the prioritized target lists using the new ZODIPIC method of calculating an exozodi surface brightness consistent with the synthetic exoEarth orbital parameters.

For a given value of $n$, the prioritized target lists are qualitatively similar, with the biggest difference occurring at the large distance limit for early type stars.  For the two values of $n$ shown, the ZODIPIC treatment of exozodi surface brightness increased yield by $\sim 7\%$.  We compared the yield calculated via both methods over a coarse grid of models and found that the ZODIPIC treatment increased yield anywhere from 0--12\%, with a median increase of $\sim5$\%.

Therefore, geometric and scattering effects appear to increase the exoEarth candidate yield by a negligible amount, and only marginally impact the prioritization of the target list.  These changes come with the caveat that one must assume a given disk geometry, most notably the scale height and radial density distribution, as well as a given scattering phase function for the dust, which certainly will not be valid for all disks.  Calculating an exozodiacal surface brightness for each synthetic exoEarth also increases the run time of the code by a few tens of percent.  In light of this, \emph{we opted not to include the effects of realistic disk geometry in our exozodi brightness calculations.}

\subsection{Impact of spectral type}

The V band surface brightness of a face-on optically thin disk with scale height $\ll1$ at circumstellar distance $r$,
\begin{equation}
\label{disk_sb_equation}
I^{\rm disk}_{\rm V}\!\left(r\right) \propto F^{\star}_{\rm V}\!\left(r\right) \tau\!\left(r\right),
\end{equation}
where $F^{\star}_{\rm V}(r)$ is the V band stellar flux at $r$ and $\tau(r)$ is the optical depth at $r$.  Assuming all stars are equally ``dusty," i.e. have equal optical depths in their habitable zones, the surface brightness should change with spectral type.  Conversely, assuming a constant surface brightness for 1 zodi of exozodiacal light necessarily implies that late type stars have exozodiacal disks that are ``dustier" than early type stars, for which no observational evidence exists.  Thus, a constant surface brightness independent of spectral type unfairly penalizes late type stars.

Here we investigate the impact of using a constant habitable zone dust optical depth opposed to a constant surface brightness.  In the previous section we showed that the disk geometry and the $1/r^2$ illumination factor within a given exozodiacal disk negligibly impact the exoEarth candidate yield and require a number of unconstrained assumptions.  Here we ignore how the exozodiacal surface brightness may change with a planet's position around a given star, and apply the surface brightness of an exozodiacal disk at the Earth-equivalent insolation distance (EEID) to all synthetic exoEarths around that star.

Using Equation \ref{disk_sb_equation}, we can express the optical depth of the zodiacal cloud at 1 AU as
\begin{equation}
\label{ez_tau_equation}
  \tau\!\left(r=1\, {\rm AU}\right) \propto I^{\rm zodiacal}_{\rm V}\!\left(r=1\, {\rm AU}\right) F^{\sun}_{\rm V}\!\left(r=1\, {\rm AU}\right).
\end{equation}
Assuming the optical depth is constant at the EEID,
\begin{equation}
  I^{\rm disk}_{\rm V}\!\left(r\right) = \frac{F^{\star}_{\rm V}\!\left(r={\rm EEID}\right)}{F^{\sun}_{\rm V}\!\left(r=1\, {\rm AU}\right)} I^{\rm zodiacal}_{\rm V}\!\left(r=1\, {\rm AU}\right).
\end{equation}
Given that EEID$=(1\, {\rm AU}) \times (L_{\star} / L_{\sun})^{0.5}$,
\begin{equation}
\label{ez_sb_equation}
  I^{\rm disk}_{\rm V}\!\left(r\right) = 10^{-0.4\left(M^{\star}_{\rm V} - M^{\sun}_{\rm V}\right)} \left(\frac{L_{\sun}}{L_{\star}}\right) I^{\rm zodiacal}_{\rm V}\!\left(r=1\, {\rm AU}\right),
\end{equation}
where $M^{\sun}_{\rm V} = 4.83$ is the absolute V band magnitude of the Sun and $M^{\star}_{\rm V}$ is the absolute V band magnitude of the star.

As discussed in Appendix \ref{zodi_appendix}, the V band surface brightness of the zodiacal cloud near 1 AU, when viewed from within the cloud, is approximately 23 mag arcsec$^{-2}$. When observed from afar, the zodiacal cloud should appear twice as bright ($\approx 22.25$ mag arcsec$^{-2}$), as we would typically observe the dust both above and below the mid-plane.  The zodiacal cloud should appear even brighter from afar for two reasons: the additional dust is observed at forward scattering angles, and the $1/r^2$ illumination factor will bias the background flux in any photometric aperture toward brighter values. Thus, we set $I^{\rm zodiacal}_{\rm V}(r=1\, {\rm AU}) = 22$ mag arcsec$^{-2}$. 

For later type stars, the V band flux drops off more rapidly than the bolometric flux; the decrease in surface brightness due to decreased V band flux outweighs the increase in surface brightness due to a more compact habitable zone.  For typical correlations between $M^{\star}_{\rm V}$ and $L_{\star}$, Equation \ref{ez_sb_equation} implies that each zodi of exozodiacal dust around a late M star is $\sim2.5$ times dimmer than around a Sun-like star in V band.

The right panels in Figure \ref{exozodi_treatment_comparison_fig} show the prioritized target lists using a constant habitable zone optical depth for $n=3$ and $n=30$ zodis of dust.  Compared to the constant surface brightness method, shown in the left panels, the yield is increased by a few percent.  The prioritized target lists also appear qualitatively similar, but on closer inspection the constant optical depth treatment increases the number of observed K and M stars by $\sim15\%$ and $\sim20\%$, respectively.

We compared the yield calculated via both methods over a coarse grid of models and found that the constant optical depth treatment increased yield anywhere from 0--17\%, with a median increase of $\sim4$\%.  The largest gains occurred for cases with small inner working angles, such that the compact habitable zones of late type stars were accessible, and large exozodi levels such that exozodi strongly dominated the background count rate.  Given that the additional run time to include this effect is negligible, \emph{we opted to use this constant optical depth exozodi treatment.}  Thus, we define one zodi of exozodiacal dust as the optical depth of the zodiacal cloud at 1 AU (see Equation \ref{ez_tau_equation}), and for a given star approximate the exozodiacal surface brightness for all synthetic exoEarths as the surface brightness at the EEID.

\end{appendices}

\bibliography{ms_v2}

\begin{deluxetable}{ccccccccccccc}
\tablewidth{0pt}
\footnotesize
\tablecaption{ExoEarth Candidate Yield, Exposure Times, \& Sensitivities \label{yield_sensitivities_table}}
\tablehead{
\colhead{$D$} & \colhead{IWA} & \colhead{$\zeta$} & \colhead{$(\zeta_{\rm floor}/\zeta)$\tablenotemark{\dagger}} & \colhead{$n$} & \colhead{$N_{\rm EC}$} & \colhead{$\langle t_{\rm expose}\rangle$} & \colhead{$\phi_D$} & \colhead{$\phi_{\rm IWA}$} & \colhead{$\phi_{\zeta}$} & \colhead{$\phi_{\zeta_{\rm floor}/\zeta}$} & \colhead{$\phi_n$}\\
\colhead{(m)} & \colhead{($\lambda/D$)} & & & \colhead{(zodis)} & & \colhead{(d)} & & & & & \\
}
\startdata
$4$ & $2$ & $10^{-10}$ & $0.40$ & $3$ & 3.80 & 2.98 & 2.07 & -1.10 & -0.20 & -0.17 & -0.23 \\
$4$ & $2$ & $10^{-10}$ & $0.40$ & $30$ & 2.00 & 6.01 & 2.09 & -0.81 & -0.10 & -0.10 & -0.37 \\
$4$ & $2$ & $10^{-10}$ & $0.06$ & $3$ & 4.13 & 3.10 & 1.96 & -0.89 & -0.08 & -0.01 & -0.24 \\
$4$ & $2$ & $10^{-10}$ & $0.06$ & $30$ & 2.11 & 5.91 & 2.00 & -0.73 & -0.02 & -0.00 & -0.38 \\
$4$ & $2$ & $10^{-9}$ & $0.40$ & $3$ & 0.66 & 6.07 & 3.15 & -3.47 & -1.13 & -1.13 & -0.00 \\
$4$ & $2$ & $10^{-9}$ & $0.40$ & $30$ & 0.64 & 8.85 & 2.88 & -2.93 & -1.03 & -0.83 & -0.04 \\
$4$ & $2$ & $10^{-9}$ & $0.06$ & $3$ & 2.70 & 4.05 & 2.00 & -1.25 & -0.34 & -0.18 & -0.11 \\
$4$ & $2$ & $10^{-9}$ & $0.06$ & $30$ & 1.76 & 6.87 & 2.05 & -0.94 & -0.19 & -0.14 & -0.27 \\
$4$ & $4$ & $10^{-10}$ & $0.40$ & $3$ & 1.03 & 3.38 & 3.14 & -3.04 & -0.77 & -0.62 & -0.02 \\
$4$ & $4$ & $10^{-10}$ & $0.40$ & $30$ & 0.82 & 9.39 & 2.63 & -2.08 & -0.43 & -0.38 & -0.18 \\
$4$ & $4$ & $10^{-10}$ & $0.06$ & $3$ & 1.89 & 6.44 & 2.24 & -1.45 & -0.13 & -0.05 & -0.16 \\
$4$ & $4$ & $10^{-10}$ & $0.06$ & $30$ & 1.10 & 11.66 & 2.16 & -1.17 & -0.04 & -0.03 & -0.32 \\
$4$ & $4$ & $10^{-9}$ & $0.40$ & $3$ & 0.06 & 0.11 & 4.01 & -3.40 & -1.80 & -1.81 & -0.00 \\
$4$ & $4$ & $10^{-9}$ & $0.40$ & $30$ & 0.06 & 0.36 & 4.00 & -3.40 & -1.80 & -1.81 & 0.00 \\
$4$ & $4$ & $10^{-9}$ & $0.06$ & $3$ & 0.62 & 4.20 & 3.23 & -3.27 & -1.00 & -0.68 & -0.00 \\
$4$ & $4$ & $10^{-9}$ & $0.06$ & $30$ & 0.59 & 9.83 & 2.79 & -2.56 & -0.75 & -0.51 & -0.07 \\
$8$ & $2$ & $10^{-10}$ & $0.40$ & $3$ & 15.10 & 0.80 & 1.96 & -0.80 & -0.15 & -0.09 & -0.22 \\
$8$ & $2$ & $10^{-10}$ & $0.40$ & $30$ & 8.05 & 1.59 & 1.89 & -0.59 & -0.08 & -0.06 & -0.36 \\
$8$ & $2$ & $10^{-10}$ & $0.06$ & $3$ & 15.64 & 0.81 & 1.89 & -0.72 & -0.09 & -0.00 & -0.22 \\
$8$ & $2$ & $10^{-10}$ & $0.06$ & $30$ & 8.25 & 1.55 & 1.85 & -0.55 & -0.03 & -0.00 & -0.37 \\
$8$ & $2$ & $10^{-9}$ & $0.40$ & $3$ & 4.93 & 1.34 & 2.77 & -2.53 & -0.90 & -0.66 & -0.04 \\
$8$ & $2$ & $10^{-9}$ & $0.40$ & $30$ & 3.96 & 2.13 & 2.49 & -1.73 & -0.62 & -0.49 & -0.16 \\
$8$ & $2$ & $10^{-9}$ & $0.06$ & $3$ & 10.22 & 1.18 & 1.88 & -0.94 & -0.28 & -0.07 & -0.09 \\
$8$ & $2$ & $10^{-9}$ & $0.06$ & $30$ & 6.91 & 1.76 & 1.86 & -0.68 & -0.17 & -0.07 & -0.26 \\
$8$ & $4$ & $10^{-10}$ & $0.40$ & $3$ & 6.86 & 1.24 & 2.31 & -1.77 & -0.46 & -0.36 & -0.10 \\
$8$ & $4$ & $10^{-10}$ & $0.40$ & $30$ & 4.60 & 2.33 & 2.15 & -1.23 & -0.23 & -0.21 & -0.27 \\
$8$ & $4$ & $10^{-10}$ & $0.06$ & $3$ & 8.58 & 1.47 & 1.94 & -1.12 & -0.13 & -0.01 & -0.15 \\
$8$ & $4$ & $10^{-10}$ & $0.06$ & $30$ & 5.06 & 2.35 & 2.03 & -0.99 & -0.04 & -0.00 & -0.33 \\
$8$ & $4$ & $10^{-9}$ & $0.40$ & $3$ & 0.66 & 11.89 & 3.15 & -3.27 & -1.13 & -1.20 & -0.00 \\
$8$ & $4$ & $10^{-9}$ & $0.40$ & $30$ & 0.66 & 9.95 & 3.15 & -3.26 & -1.12 & -1.16 & -0.00 \\
$8$ & $4$ & $10^{-9}$ & $0.06$ & $3$ & 4.41 & 2.00 & 2.22 & -1.81 & -0.59 & -0.35 & -0.03 \\
$8$ & $4$ & $10^{-9}$ & $0.06$ & $30$ & 3.60 & 2.74 & 2.20 & -1.48 & -0.40 & -0.29 & -0.16 \\
$12$ & $2$ & $10^{-10}$ & $0.40$ & $3$ & 33.08 & 0.38 & 1.87 & -0.65 & -0.13 & -0.05 & -0.22 \\
$12$ & $2$ & $10^{-10}$ & $0.40$ & $30$ & 17.21 & 0.74 & 1.85 & -0.52 & -0.06 & -0.04 & -0.37 \\
$12$ & $2$ & $10^{-10}$ & $0.06$ & $3$ & 33.54 & 0.38 & 1.84 & -0.62 & -0.10 & -0.00 & -0.22 \\
$12$ & $2$ & $10^{-10}$ & $0.06$ & $30$ & 17.43 & 0.72 & 1.82 & -0.50 & -0.03 & -0.00 & -0.38 \\
$12$ & $2$ & $10^{-9}$ & $0.40$ & $3$ & 14.22 & 0.57 & 2.44 & -1.80 & -0.70 & -0.47 & -0.07 \\
$12$ & $2$ & $10^{-9}$ & $0.40$ & $30$ & 10.30 & 0.99 & 2.18 & -1.20 & -0.46 & -0.36 & -0.22 \\
$12$ & $2$ & $10^{-9}$ & $0.06$ & $3$ & 21.61 & 0.56 & 1.79 & -0.81 & -0.27 & -0.03 & -0.09 \\
$12$ & $2$ & $10^{-9}$ & $0.06$ & $30$ & 14.64 & 0.84 & 1.82 & -0.60 & -0.15 & -0.04 & -0.26 \\
$12$ & $4$ & $10^{-10}$ & $0.40$ & $3$ & 17.09 & 0.60 & 2.21 & -1.41 & -0.31 & -0.23 & -0.15 \\
$12$ & $4$ & $10^{-10}$ & $0.40$ & $30$ & 10.36 & 1.13 & 1.96 & -0.98 & -0.17 & -0.14 & -0.29 \\
$12$ & $4$ & $10^{-10}$ & $0.06$ & $3$ & 18.57 & 0.63 & 1.98 & -1.13 & -0.14 & -0.00 & -0.16 \\
$12$ & $4$ & $10^{-10}$ & $0.06$ & $30$ & 10.90 & 1.13 & 1.84 & -0.85 & -0.04 & -0.00 & -0.31 \\
$12$ & $4$ & $10^{-9}$ & $0.40$ & $3$ & 2.19 & 2.65 & 3.02 & -3.20 & -1.08 & -1.14 & -0.00 \\
$12$ & $4$ & $10^{-9}$ & $0.40$ & $30$ & 2.19 & 3.26 & 3.00 & -3.10 & -1.05 & -1.00 & -0.01 \\
$12$ & $4$ & $10^{-9}$ & $0.06$ & $3$ & 10.22 & 1.00 & 2.07 & -1.44 & -0.42 & -0.21 & -0.04 \\
$12$ & $4$ & $10^{-9}$ & $0.06$ & $30$ & 8.15 & 1.36 & 1.95 & -1.14 & -0.30 & -0.18 & -0.16 \\
$16$ & $2$ & $10^{-10}$ & $0.40$ & $3$ & 56.45 & 0.22 & 1.82 & -0.57 & -0.12 & -0.03 & -0.21 \\
$16$ & $2$ & $10^{-10}$ & $0.40$ & $30$ & 29.17 & 0.43 & 1.81 & -0.45 & -0.05 & -0.03 & -0.38 \\
$16$ & $2$ & $10^{-10}$ & $0.06$ & $3$ & 56.76 & 0.22 & 1.80 & -0.56 & -0.10 & -0.00 & -0.21 \\
$16$ & $2$ & $10^{-10}$ & $0.06$ & $30$ & 29.32 & 0.43 & 1.79 & -0.44 & -0.03 & -0.00 & -0.38 \\
$16$ & $2$ & $10^{-9}$ & $0.40$ & $3$ & 27.63 & 0.35 & 2.12 & -1.33 & -0.59 & -0.37 & -0.08 \\
$16$ & $2$ & $10^{-9}$ & $0.40$ & $30$ & 18.91 & 0.56 & 2.04 & -0.93 & -0.38 & -0.29 & -0.25 \\
$16$ & $2$ & $10^{-9}$ & $0.06$ & $3$ & 36.06 & 0.33 & 1.74 & -0.73 & -0.27 & -0.02 & -0.09 \\
$16$ & $2$ & $10^{-9}$ & $0.06$ & $30$ & 24.48 & 0.50 & 1.75 & -0.52 & -0.14 & -0.03 & -0.26 \\
$16$ & $4$ & $10^{-10}$ & $0.40$ & $3$ & 31.61 & 0.36 & 2.06 & -1.18 & -0.24 & -0.15 & -0.15 \\
$16$ & $4$ & $10^{-10}$ & $0.40$ & $30$ & 18.50 & 0.62 & 2.02 & -0.94 & -0.11 & -0.09 & -0.33 \\
$16$ & $4$ & $10^{-10}$ & $0.06$ & $3$ & 32.84 & 0.36 & 1.94 & -1.04 & -0.14 & -0.00 & -0.15 \\
$16$ & $4$ & $10^{-10}$ & $0.06$ & $30$ & 18.91 & 0.61 & 1.95 & -0.88 & -0.04 & -0.00 & -0.34 \\
$16$ & $4$ & $10^{-9}$ & $0.40$ & $3$ & 5.25 & 0.85 & 3.05 & -3.21 & -1.09 & -1.05 & -0.00 \\
$16$ & $4$ & $10^{-9}$ & $0.40$ & $30$ & 5.17 & 1.05 & 2.94 & -2.93 & -1.00 & -0.86 & -0.02 \\
$16$ & $4$ & $10^{-9}$ & $0.06$ & $3$ & 18.39 & 0.61 & 1.95 & -1.26 & -0.35 & -0.13 & -0.04 \\
$16$ & $4$ & $10^{-9}$ & $0.06$ & $30$ & 14.45 & 0.76 & 1.99 & -1.07 & -0.24 & -0.11 & -0.18 \\

\enddata
\vspace{-0.25in}
 \tablecomments{ExoEarth candidate yield $N_{\rm EC}$, completeness-weighted average exposure time $\langle \tau \rangle$, and sensitivities $\phi_x$ for all mission parameters investigated (only a sparse sampling of phase space is listed).  Sensitivity $\phi_x$ to changes in parameter $x$ is equivalent to the percent change in yield per percent change in $x$, and the yield roughly scales locally as $N_{\rm EC} \propto x^{\phi_x}$.  Parameters listed are telescope diameter ($D$), coronagraph inner working angle (IWA), coronagraph raw contrast ($\zeta$), contrast floor to raw contrast ratio ($\zeta_{\rm floor}/\zeta$), and exozodi level ($n$).}
\tablenotetext{\dagger}{$(\zeta_{\rm floor}/\zeta)=(10^{-0.4\Delta{\rm mag_{floor}}})/\zeta$.  $\Delta{\rm mag_{floor}}$ defines the dimmest point source detectable at the chosen S/N, such that $\zeta_{\rm floor}$ is the flux ratio of the dimmest detectable point source at the chosen S/N.  For this work, we set S/N $=10$.}
\end{deluxetable}

\end{document}